\documentclass[conference]{IEEEtran}
\IEEEoverridecommandlockouts
\usepackage{mathrsfs}
\usepackage[lined,boxed,commentsnumbered,ruled]{algorithm2e}
\usepackage{algorithmic}
\usepackage{array}
\usepackage{subfigure}
\usepackage{url}
\usepackage{graphicx}
\usepackage{indentfirst}
\usepackage{adjustbox}
\usepackage[perpage,symbol*]{footmisc}
\usepackage{array}
\usepackage{subfigure}
\usepackage{cite}
\usepackage{amsmath,amssymb,amsfonts}
\usepackage{algorithmic}
\usepackage{graphicx}
\usepackage{textcomp}
\usepackage{xcolor}
\def\BibTeX{{\rm B\kern-.05em{\sc i\kern-.025em b}\kern-.08em
    T\kern-.1667em\lower.7ex\hbox{E}\kern-.125emX}}

\newtheorem{proposition}{Proposition}

\newtheorem{observation}{Observation}
\newtheorem{definition}{Definition}
\newtheorem{example}{Example}

\newtheorem{theorem}{Theorem}

\newcommand{\nop}[1]{}

\begin{document}

\title{Finding Route Hotspots in Large Labeled Networks}

\author{\IEEEauthorblockN{Mingtao Lei\IEEEauthorrefmark{2}, Xi Zhang\IEEEauthorrefmark{2}\thanks{\IEEEauthorrefmark{1}Corresponding author.}\IEEEauthorrefmark{1}, Lingyang Chu\IEEEauthorrefmark{3},  Zhefeng Wang\IEEEauthorrefmark{4}, Philip S. Yu\IEEEauthorrefmark{6}, Binxing Fang\IEEEauthorrefmark{2}} 
\IEEEauthorblockA{\IEEEauthorrefmark{2}Key Laboratory of Trustworthy Distributed Computing and Service (BUPT), Ministry of Education, \\
Beijing University of Posts and Telecommunications, Beijing, China} 
\IEEEauthorblockA{\IEEEauthorrefmark{3}Huawei Technologies Canada, Burnaby, Canada.}
\IEEEauthorblockA{\IEEEauthorrefmark{4}Huawei Technologies, Hangzhou, China}
\IEEEauthorblockA{\IEEEauthorrefmark{6}Department of Computer Science, University of Illinois at Chicago, IL, USA}
\{leimingtao, zhangx, fangbx\}@bupt.edu.cn, lingyang.chu1@huawei.com, wangzhefeng@huawei.com, psyu@uic.edu
}
\maketitle

\begin{abstract}
In many advanced network analysis applications, like social networks, e-commerce, and network security, hotspots are generally considered as a group of vertices that are tightly connected owing to the similar characteristics, such as common habits and location proximity. In this paper, we investigate the formation of hotspots from an alternative perspective that considers the routes along the network paths as the auxiliary information, and attempt to find the route hotspots in large labeled networks. A \textit{route hotspot} is a cohesive subgraph that is covered by a set of routes, and these routes correspond to the same sequential pattern consisting of vertices' labels. To the best of our knowledge, the problem of \textit{Finding Route Hotspots in Large Labeled Networks} has not been tackled in the literature. However, it is challenging as counting the number of hotspots in a network is \#P-hard. Inspired by the observation that the sizes of hotspots decrease with the increasing lengths of patterns, we prove several anti-monotonicity properties of hotspots, and then develop a scalable algorithm called FastRH that can use these properties to effectively prune the patterns that cannot form any hotspots. In addition, to avoid the duplicate computation overhead, we judiciously design an effective index structure called \textit{RH-Index} for storing the hotspot and pattern information collectively, which also enables incremental updating and efficient query processing. Our experimental results on real-world datasets clearly demonstrate the effectiveness and scalability of our proposed methods.
\end{abstract}


\section{Introduction}\label{sec:introduction}
Finding communities from real-world networks has gained significant interests, which enjoys various applications involving social networks, e-commerce, and network security. Conventional community detection methods detect communities as groups of vertices that are densely interconnected, which focus on the graph structures of communities. More often than not, besides the topology information, the valuable co-occurrence information~\cite{Chu2019} and route information~\cite{Shang2018} may also help to find meaningful communities. 

\begin{figure}[htb]
\centering
	\subfigure[]{\includegraphics[scale=0.25]{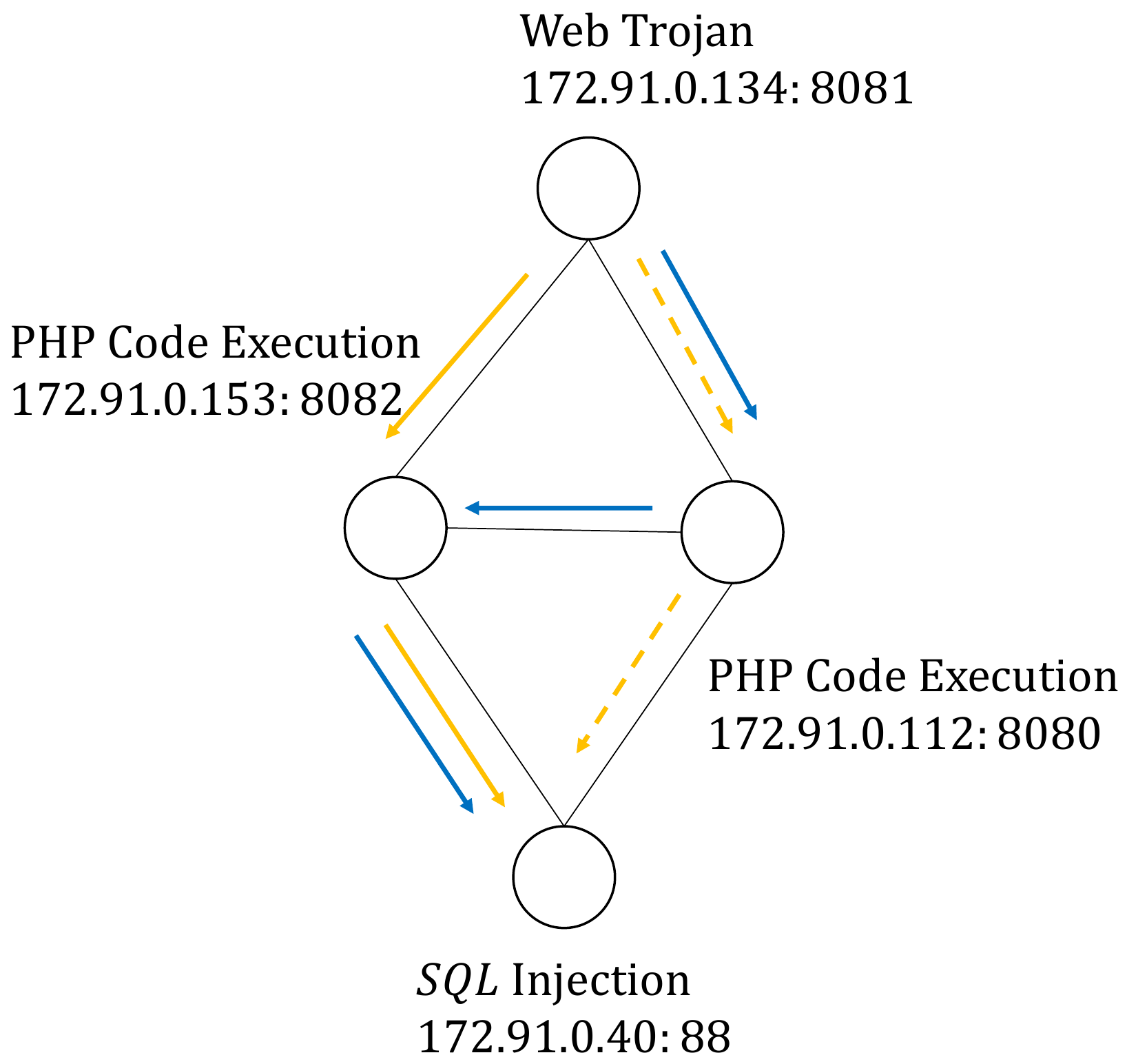}}
	\subfigure[]{\includegraphics[scale=0.25]{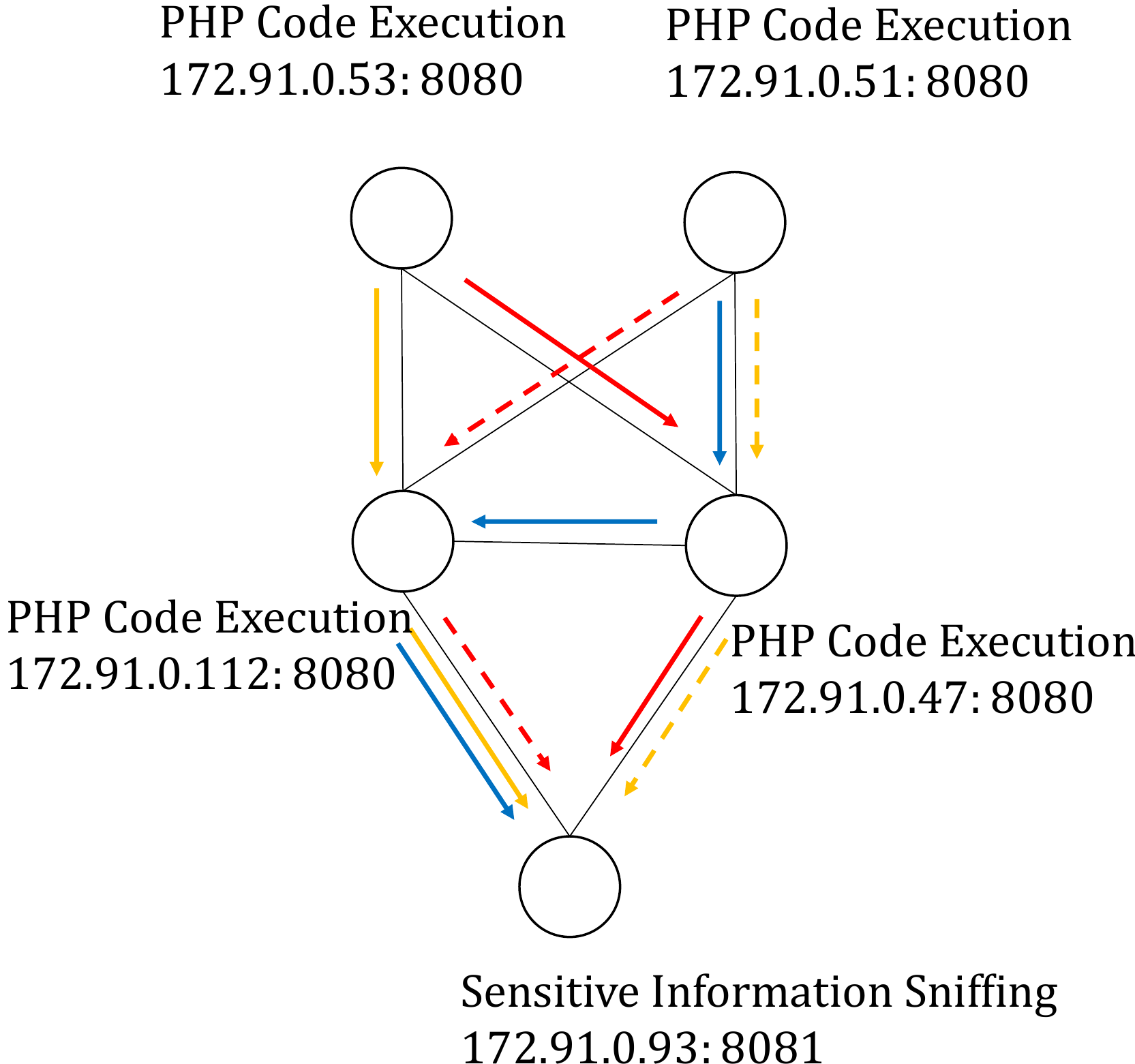}}			
\caption{Two examples of network security, where each vertex corresponds to a tuple containing an IP address, a port and a specific vulnerability. The colored arrow lines on edges represent routes, where a route is a sequence of consecutive edges, incidenting a chain of attacks. Fig. 1(a) shows that hackers frequently first exploit the vulnerabilities \textit{Web Trojan} and \textit{PHP Code Execution}, and then exploit the vulnerability \textit{SQL Injection}. Fig. 1(b) indicates that hackers frequently first exploit the vulnerability \textit{PHP Code Execution} on specific systems, then exploit the same kind of vulnerability (i.e., \textit{PHP Code Execution}) on some other systems, and further exploit the vulnerability \textit{Sensitive Information Sniffing}. These two examples show two types of successful attacking paths in network security.}
\label{fig:security}
\end{figure}

The route on the graph, which is a sequence of consecutive edges, is a natural descriptive model for many real-world networks. The \textit{route} defined here is similar to that defined in~\cite{Dai2015}, which is also a sequence of edges and the consecutive edges must share a vertex. Examples in network security are shown in Figure~\ref{fig:security}, where each vertex corresponds to a tuple containing an IP address, a port and a specific vulnerability. And each edge indicates that the two connected systems are incident in the network. Hackers may find a sequence of exploits (i.e., a route) along the network path that can enable a successful attack. For example, the routes marked by the colored arrow lines indicate that some hackers frequently first exploit the vulnerabilities \textit{Web Trojan} and \textit{PHP Code Execution}, and then exploit the vulnerability \textit{SQL Injection} (Fig. 1(a)), while some other hackers frequently first exploit the vulnerability \textit{PHP Code Execution} on some specific systems, then exploit the same kind of vulnerability (i.e., \textit{PHP Code Execution}) on other systems, and further exploit the vulnerability \textit{Sensitive Information Sniffing} (Fig. 1(b)). Another example lies in the anti-money laundering analysis~\cite{Lopez2016}, where each vertex is a user of a bank and an edge represents their money transfer relationship. Here a route can denote a sequence of transfer behaviours among a set of users. The routes may also exist in the collaboration networks, such as DBLP~\cite{DBLP}, where each vertex is an author and an edge represents their collaboration relationship. In addition to collaboration relationships, we also consider the authors' citation relationships to build the citation sequence that is used as the route. Specifically, for every two consecutive collaborative authors A and B in the route, B should have cited a paper published by A. Such routes can indicate the trends of their research interests.

Informally, given a graph as well as routes on it, it is interesting to discover dense components which are covered by groups of routes that share the same patterns implying common behaviours. In the examples of network security that are shown in Figure~\ref{fig:security}, the detection of these components can facilitate in both identifying the security vulnerable areas in the networks and extracting the features of hacker behaviours, and these features can be used for hacker identification. In the example of the anti-money laundering analysis, the frequent money transfer behaviours among a specific group of users could be suspicious, and the detection of such users may indicate money laundering. In the third example given above, the detection can discover groups of closely cooperated scholars who hold the same continuous research interests in the collaboration networks. We call such dense components \textit{route hotspots} and aim to discover the route hotspots in an efficient manner.

We can also understand the route hotspots detection problem from the perspective of the patterns. It would also be interesting to find the same patterns that are shared by a specific dense component. Although a few previous studies have attempted to mine frequent patterns from a large graph, a pattern that is frequent in the whole graph is not necessarily frequent in a dense component of the graph. Thus, the detection of route hotspots can also help to find meaningful patterns regarding to a specific component of the graph.

However, effectively \textit{\textbf{finding route hotspots in large labeled networks}} is a non-trivial problem. It is challenging on how to define the route hotspot appropriately and how to develop an efficient solution for route hotspot finding.

\subsection{Challenge 1: Suitable Model}
We first review some related models defined in the previous work and then conclude the key attributes that are required to be possessed by our model. 

\begin{figure}[t]
\centering
	\includegraphics[scale=0.3]{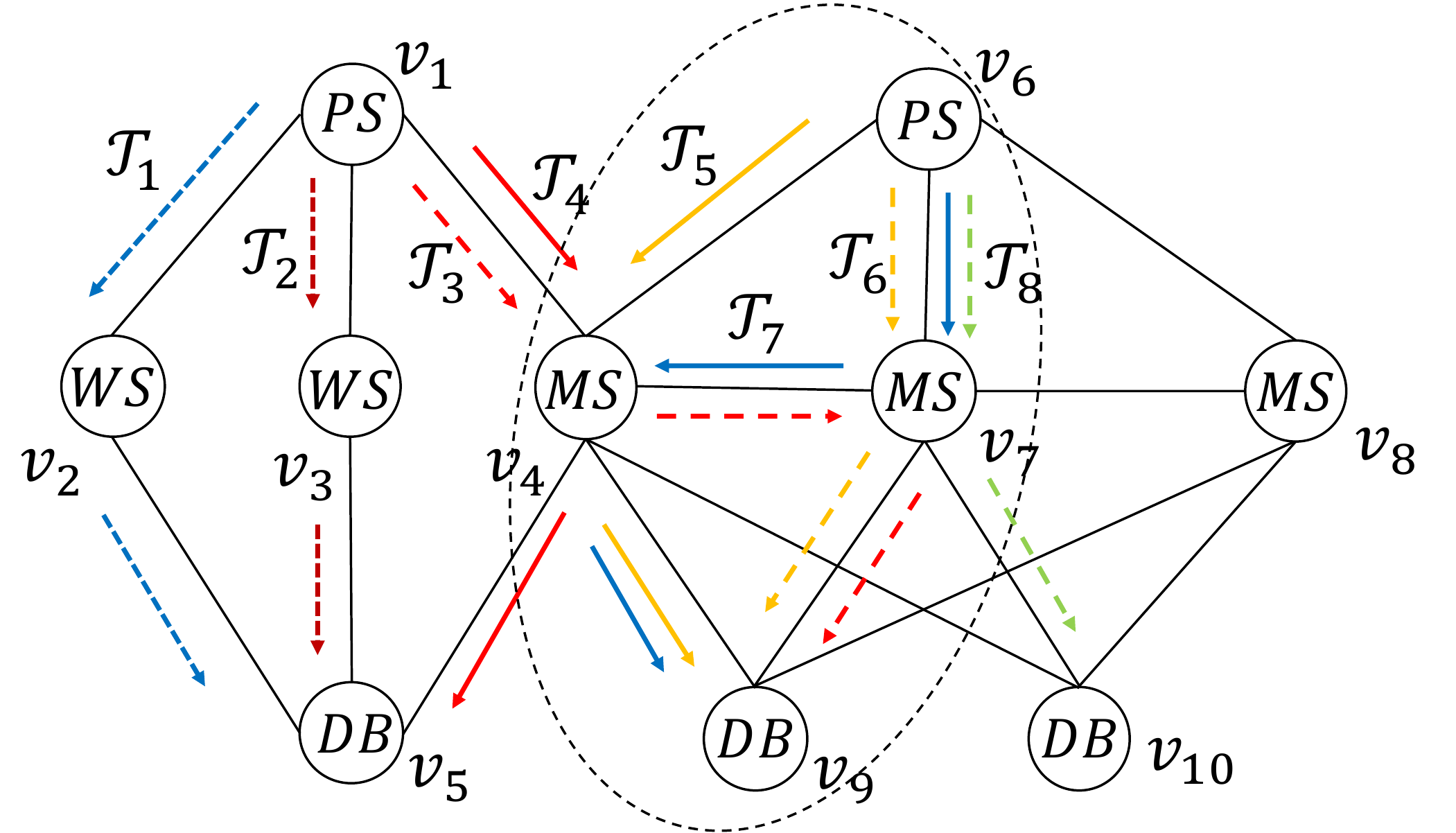}
\caption{An example network, where each vertex corresponds to a specific label, including DB, WS, MS and PS respectively. The colored arrow lines on edges represent routes, where a route $\mathcal{T}$ is a sequence of consecutive edges. The dashed ellipse represents a route hotspot (i.e., the induced subgraph $H$ of $\{v_4, v_6, v_7, v_9\}$) that can be marked by the sequential pattern $\langle PS, MS, DB \rangle$, which is covered by qualified routes that form this hotspot.}
\label{fig:ex}
\end{figure}

On one hand, various community models have been proposed to describe dense subgraphs, such as $k$-core~\cite{Peng2018} and $k$-truss~\cite{Wang2012}. But these dense subgraphs cannot be adopted to model route hotspots as they mainly focus on the graph structures but ignore how routes perform. To better introduce the model, we give a more specific example of network security in Figure~\ref{fig:ex}, where each vertex in the graph corresponds to a system vulnerability and is labeled as its type, such as DB (the database), WS (the website system), MS (the management system) and PS (the portal system). The induced subgraph of $\{v_4, v_7, v_8, v_9, v_{10}\}$ is a $4$-truss. However, there are no routes traversing the edges $(v_7, v_8)$, $(v_8, v_9)$ and $(v_8, v_{10})$ in this subgraph. A \textit{coverage} relationship that requires the subgraph is covered by a group of routes may make more sense.

On the other hand, there also exist some concepts with the goal of discovering a group of routes that move together, such as \textit{convoy}~\cite{Jeung2008} and \textit{trajectory gathering pattern}~\cite{Zheng2013}. These concepts can be distinguished based on how the ``group" is defined. However, using such group patterns to model route hotspots is limited by the following two issues. First, although these approaches are able to identify a group of objects moving together, they only consider spatial proximity for clustering a group of objects rather than more refined and meaningful interconnection relationships between pair-wised objects. Moreover, the group of objects they find may not be densely interconnected and thus are not suitable for our problem settings. For example, as shown in Figure~\ref{fig:ex}, the subgraph of $\{v_1, v_2, v_3, v_4, v_5\}$ is obtained by spatial clustering and is covered by $\{\mathcal{T}_1, \mathcal{T}_2, \mathcal{T}_4\}$, but it is not a dense structure. Second, the output of these approaches may be varied by using different distance measures to obtain clustering results. In contrast, our proposal working with $k$-truss to indicate dense groups is independent of distance measures. 

With the insights given above, we regard a route hotspot as a cohesive subgraph covered by a set of routes containing the same sequential pattern. An appropriate model of route hotspot should possess the following key attributes.

\begin{enumerate}
	\item (Cohesiveness) The value of the cohesiveness function that measures the structure of route hotspot is high.

	\item (Participation) The route hotspot should contain at least $min\_sup$ subroutes (a subroute is one of the consecutive portions of the full route).	
	
	\item (Coverage) These subroutes contained by the route hotspot can cover the route hotspot.
	
	\item (Correlation) Each route contained in the route hotspot should contain the same sequential pattern.
	
\end{enumerate}

The cohesiveness condition is straightforward as hotspots are supposed to be dense subgraphs. Here we use $k$-truss as it has a stronger guarantee on cohesive structure than $k$-core. Furthermore, computing $k$-truss subgraphs takes low computational cost, that is, polynomial time~\cite{Wang2012}. The participation and coverage conditions are also straightforward as the hotspot is defined on the graph and route set. The correlation condition shows that, in the route hotspot, routes should contain the same sequential pattern, which is a sequence of consecutive labels, and thus we can mark the hotspot by this pattern. As shown in Figure~\ref{fig:ex}, let $k=3$ and $min\_sup=3$. The induced subgraph $H$ of $\{v_4, v_6, v_7, v_9\}$ is a $4$-truss, which contains $\mathcal{T}_5$, $\mathcal{T}_6$, $\mathcal{T}_7$ and one portion of $\mathcal{T}_8$, where each route contain the same sequential pattern $\langle PS, MS \rangle$ and these routes can cover $H$. Therefore, $H$ is a route hotspot for $k=3$ and pattern $\langle PS, MS \rangle$. We will give more formal definitions in Section~\ref{sec:prob}.

\subsection{Challenge 2: Efficient Detection Algorithm}
It is natural to ask whether we can apply or extend the existing algorithms for sequential pattern mining, community detection, or group pattern discovery to discover route hotspots? Unfortunately, the answer is no due to the following challenges.

First, the conventional sequential pattern mining methods, like PrefixSpan~\cite{Pei2001}, can only find the patterns that exist in the routes and describe how patterns are located on the graph. They don't consider dense graph structures.

Second, the routes may contain an exponential number of patterns. Since the conventional community detection methods, such as BULK~\cite{Huang2017}, can only detect route hotspots of one pattern at a time, it is computationally intractable to use those methods for each of the exponential number of patterns.

Third, the group pattern discovery methods, like CuTS~\cite{Jeung2008}, are not designed to process network topology information and thus the group of vertices they detect may not be densely connected from the perspective of the graph structure such as $k$-truss.

Last but not least, it is of practical importance to provide a fast query service to answer user queries for interesting route hotspots in real time. As the number of route hotspots is usually huge in a large labeled network, enumerating and indexing all of them efficiently is extremely challenging. 

To tackle these challenges, we make the following contributions in this work.

First, to the best of our knowledge, we are the first to detect the route hotspots in the large labeled networks. We prove that counting the number of route hotspots is \#P-hard and propose to solve the problem by exploring the graph structure and sequential patterns simultaneously.

Second, to reduce the computational cost of the straightforward greedy algorithms, we first prove the \textit{pattern anti-monotonicity} and the \textit{hotspot anti-monotonicity} properties on route hotspots, and then develop an efficient algorithm called \textit{FastRH} via applying these properties. FastRH can also be parallelized since the detection procedure for different patterns is independent, and thus can be speeded up. 

Third, to store the huge number of decomposed route hotspots, we design a tree index structure called \textit{RH-Index}, which can efficiently index more than $8.2 \times 10^5$ hotspots for $6.7 \times 10^5$ patterns by using 2.3 hours and 39.9 GB main memory on a commodity PC with a large main memory as shown in Section~\ref{sec:experiments} on CN dataset. We also develop an efficient querying method that takes only 0.12 seconds on average for each query pattern.

Fourth, we report an extensive experimental study on real-world datasets. The results clearly show that our method is accurate, efficient and scalable in finding and indexing route hotspots.

The rest of the paper is organized as follows. We review the related work in Section~\ref{sec:relatedwork} and formulate the problem in Section~\ref{sec:prob}. In Section~\ref{sec:simpleAl} and~\ref{sec:novelAl}, we present the greedy algorithm and the improved solution. The route hotspot indexing is introduced in Section~\ref{sec:index}. We report the evaluation results in Section~\ref{sec:experiments}, and conclude the paper in Section~\ref{sec:con}. 

\section{Related Work}
\label{sec:relatedwork}
In this section, we review the related work, which can be categorized into sequential pattern mining, $k$-truss based community detection and group pattern discovery in spatial-temporal data.

\subsection{Sequential pattern mining}
\label{subsec:patternMining}
Sequential pattern mining, which is first introduced by Agrawal and Srikant in \cite{Agrawal1995}, is a well-studied topic in data mining. Given a set of sequences, the algorithms of sequential pattern mining, like FreeSpan \cite{Pei2000} and PrefixSpan \cite{Pei2001}, are proposed to find all frequent subsequences. 

To deal with the large scale of data, approximated methods \cite{Riondato2014,Riondato2015} and distributed methods \cite{Li2008,Ge2016} are proposed. Riondato~et~al.~\cite{Riondato2012} apply \textit{Vapnik-Chervonenkis (VC) dimension} to provide tight bounds on the sample size, which is linearly dependent on the VC-dimension of a range space associated with the dataset to be mined. Progressive sampling methods~\cite{Riondato2015} start from small samples and then increment the sampling sizes gradually until meeting the stopping conditions or hitting the upper bound. Li~et~al.~\cite{Li2008} parallelize the FP-Growth on distributed machines in order to reduce memory use and computational cost. Ge~et~al.~\cite{Ge2016} develop a memory-efficient distributed dynamic programming approach to mine probabilistic frequent sequential patterns. A comprehensive review for this topic can be found in \cite{DP2007}.

However, these approaches are not designed for graph data, and thus cannot be adopted for our problem.

\subsection{$k$-truss based community detection}
\label{subsec:communityDetcetion}
As there are various definitions on communities, here we mainly focus on the previous studies that are related to $k$-truss. Recently, Zhang~et al.~\cite{Zhang2017} propose to find interesting $(k, r)$-cores, where each vertex in a $(k, r)$-core connects to at least $k$ other vertices and the similarity between any two vertices of the $(k, r)$-core is bounded by a similarity threshold $r$. However, a connected $k$-core is not guaranteed to be 2-edge-connected, while $k$-truss is more rigorous than $k$-core owing to its triangle-based definition. Thus, $k$-truss based community models are proposed.

Huang~et al.~\cite{Huang2014} propose to search $k$-truss communities in large-scale networks. To support the efficient search of $k$-truss communities, they design a compact index, which can be updated incrementally under a dynamic graph setting. They also extend the $k$-truss community detection models to probabilistic graphs~\cite{Huang2016} and attributed graphs~\cite{Huang2017}. In particular, for probabilistic graphs, to meet the requirements in real-world networks, they propose the notion of $(k, \gamma)$-truss, where the probability of each edge contained in at least $(k-2)$ triangles is at least $\gamma$. For attributed graphs, motivated by balancing the attribute homogeneity and coverage, they propose the attributed truss community model and define the $(k ,d)$-truss. The proposed $(k ,d)$-truss is a connected $k$-truss containing all query nodes and each node of the $(k ,d)$-truss has a distance no larger than $d$ from every query node. Such models are proven to be effective in retrieving meaningful communities in probabilistic graphs and attributed graphs.

In contrast to previous studies, our hotspot is uniquely identified by a pattern, which is not touched by the above methods. Thus, these methods cannot be applied straightforwardly to detect route hotspots.

\subsection{Group pattern discovery in spatio-temporal data}
\label{subsec:communityDetcetion}
Discovering groups of objects that travel together in spatio-temporal data is called \textit{group pattern discovery}~\cite{Zheng2013} and has enjoyed many applications, such as movement behaviour analysis~\cite{Shang2018}. Vieira~et~al.~\cite{Vieira2009} propose to discover \textit{flocks}, where a flock is a group of objects that travel together within a disc of some user-specified size for at least $k$ consecutive timestamps. To deal with the so-called lossy-flock problem, the notion of \textit{convoy}~\cite{Jeung2008} is proposed, where the convoy is defined to have multiple objects that are densely connected with respect to a specific distance during $k$ consecutive time points. Subsequently, to describe the common behaviours in dense groups, the trajectory gathering pattern~\cite{Zheng2013} is proposed, where a gathering is regarded as a dense and continuing group of individuals. Zhang~et~al.~\cite{Zhang2016} propose GMOVE, to model human mobility from massive geo-tagged social media, by grouping the users that share similar moving behaviors.

However, these methods only consider spatial proximity for clustering a group of objects rather than more refined and meaningful interconnection relationships between
paired objects. Thus there is no straightforward way to apply these methods on route hotspots detection.

\section{Problem Formulation}
\label{sec:prob}
In this section, we first define \textit{route hotspot}, and then introduce the route hotspot finding problem.
\begin{table}[t]
\caption{Frequently used notations.}
\centering
\begin{adjustbox}{max width=\columnwidth}
\label{tab:notations}
	\begin{tabular}{|c|l|}
	\hline
	Notation & \multicolumn{1}{|c|}{Description}\\
	\hline
	$G(V, E, la)$ & The labeled undirected graph. \\
	\hline
	$\Pi$ & The union of labels in $G$. \\
	\hline
	$\mathcal{D}$, $\mathcal{D}_p$  & The route set and the subset of $\mathcal{D}_p$, where each route contains $p$.\\ 
	\hline
	$min\_sup$ & The user defined positive integer parameter. \\
	\hline
	$\mathcal{T}$, $t$ & The route and the subroute of $\mathcal{T}$. \\ 
	\hline
	$s$ & The sequence. \\ 
	\hline
	$p$ & The sequential pattern. \\ 
	\hline
	$H_p(k)$ & The route hotspot with fixed $p$ and $k$. \\
	\hline
	$n_p$ & The node of RH-Index for pattern $p$. \\
	\hline
	$\mathbb{L}_p$, $\mathbb{T}_p$, $\mathbb{E}_p$ & The linked item, routes, and edges of $n_p$. \\
	\hline
\end{tabular}
\end{adjustbox}
\end{table}

\subsection{Graph and $k$-truss}

We consider an undirected \textit{graph} $G=(V, E, la)$, where (1) $V$ is a finite set of vertices, (2) $E =\{(u,v) \mid u, v \in V\} \subseteq V \times V$ is a set of edges; (3) $la$ is a labeling function that maps each vertex $v$ to the unique label $L_v=la(v)$ in $\Pi$. $\Pi$ denotes the set of labels. We denote the set of neighbors of $v$ as $N(v)$, i.e., $N(v)=\{u \in V|(v, u) \in E\}$. The degree of $v$ is denoted by $d(v)=|N(v)|$. 

A \textit{triangle} in $G$ is a cycle of length 3. We denote a triangle as $\bigtriangleup_{uvw}$, where $u, v, w \in V$ are the three vertices on the cycle. The \textit{support of an edge} $e(u, v)$, denoted by $sup(e, G)$, is defined as $|\{\bigtriangleup_{uvw}: w \in V\}|$. We denote $e \in \bigtriangleup$, if $e$ is an edge of $\bigtriangleup$. A \textit{$k$-truss} $H(V_H, E_H)$ is a subgraph of $G$, such that $\forall e \in E_{H}, sup(e, H) \geq (k-2)$. $H$ is said to be a \textit{maximal} $k$-truss, if there does not exist another $k$-truss that can contain it.

\begin{figure}[t]
\centering
\begin{adjustbox}{max width=\columnwidth}	
	\begin{tabular}{|l|l|}
	\hline
	$\mathcal{T}_1=(1,\langle (v_1, v_2), (v_2, v_5) \rangle)$ \\
	$\mathcal{T}_2=(2,\langle (v_1, v_3), (v_3, v_5) \rangle)$ \\ 
	$\mathcal{T}_3=(3,\langle (v_1, v_4), (v_4, v_7), (v_7, v_9) \rangle)$ \\
	$\mathcal{T}_4=(4,\langle (v_1, v_4), (v_4, v_5) \rangle)$ \\ 
	$\mathcal{T}_5=(5,\langle (v_6, v_4), (v_4, v_6) \rangle)$ \\
	$\mathcal{T}_6=(6,\langle (v_6, v_7), (v_7, v_9) \rangle)$ \\ 
	$\mathcal{T}_7=(7,\langle (v_6, v_7), (v_7, v_4), (v_4, v_9) \rangle)$ \\
	 $\mathcal{T}_8=(8,\langle (v_6, v_7), (v_7, v_{10}) \rangle)$ \\ 
	\hline
	\end{tabular}
	\end{adjustbox}
\caption{A route set $\mathcal{D}$ of Figure~\ref{fig:ex}, where $\mathcal{T}_1$ is a route with $tid=1$, and $\langle PS, WS, DB \rangle$ is a sequence induced by $\mathcal{T}_1$.}
\label{fig:route}
\end{figure}

\subsection{Route on Graph}

Let $es=\langle e_1, \ldots, e_{h-1} \rangle$ be a sequence of edges, where $e_i=e(v_i, v_{i+1}) \in E$ ($1 \leq i < h$). A \textit{route} is a tuple $\mathcal{T}=(tid, es)$, where \textit{tid} is a unique route-id. We say that $e(v_i, v_{i+1})$ is contained by $\mathcal{T}$, denoted by $e(v_i, v_{i+1}) \in \mathcal{T}$. A route set $\mathcal{D}$ is a set of routes, and $|\mathcal{D}|$ denotes the number of routes. The \textit{length} of the route is $h-1$, the number of edges. 

A \textit{subroute} $t$ of $\mathcal{T}$, denoted by $t \in \mathcal{T}$, is one of the consecutive portions of $\mathcal{T}$. For example, $\langle (v_1, v_2), (v_2, v_3) \rangle$ and $\langle (v_2, v_3), (v_3, v_4) \rangle$ are subroutes of $\langle (v_1, v_2), (v_2, v_3), (v_3, v_4) \rangle$.

Given a subgraph $H=(V_H, E_H)$ of $G$, $\mathcal{T}$ is said to be \textit{route containment} with $H$, denoted by $\mathcal{T} \in H$, if $\forall e(u, v) \in \mathcal{T}$, such that $e \in E_H$. For example, given the subgraph $H$ of $G$ in Figure~\ref{fig:ex}, where $H$ is a vertex-induced subgraph induced by $\{v_4, v_6, v_7, v_9\}$, $\mathcal{T}_7$ in Figure~\ref{fig:route} is route containment with $H$ as each edge in $\mathcal{T}$ is an edge of $H$. Similarly, the subroute $\langle (v_6, v_7), (v_7, v_4) \rangle$ of $\mathcal{T}_7$ is also route containment with $H$. $\mathcal{D}$ is said to be \textit{route set containment} with $H$, if $\forall \mathcal{T} \in \mathcal{D}$, such that $\mathcal{T} \in H$, denoted by $\mathcal{D} \Subset H$. For example, given $H_1=(V_{H_1}, E_{H_1})$ induced from $\{v_1, v_2, v_3, v_5\}$ in Figure~\ref{fig:ex} and $\{\mathcal{T}_1, \mathcal{T}_2\}$ in Figure~\ref{fig:route}, $H_1$ contains not only the individual routes $\mathcal{T}_1$, $\mathcal{T}_2$, but also their route set $\{\mathcal{T}_1, \mathcal{T}_2\}$. 

$\mathcal{D}$ is said to \textit{cover} $H$, denoted by $H \boxdot \mathcal{D}$, if each edge of $H$ exists in at least one route of $\mathcal{D}$. Note that if $\mathcal{D}$ covers $H$, $H$ must contain the route set $\mathcal{D}$, but not vice versa. Continued with the above example, $H$  is covered by $\{\mathcal{T}_1, \mathcal{T}_2\}$, but neither $\mathcal{T}_1$ nor $\mathcal{T}_2$ can cover $H$ on its own.

\subsection{Sequence and Pattern}
We are interested in sequential patterns carried by routes in $\mathcal{D}$. Formally, a \textit{sequence} $s=\langle L_1, \ldots, L_h \rangle$ is a sequence of labels, which is \textit{induced} by $\mathcal{T}$, where $L_i=L_{v_i}$ for $1 \leq i \leq h$. $L_i$ is called the $i$-th \textit{item} of $s$. The length of $s$ is $len(s)=h$.

A sequence $p=\langle A_1, \dots, A_l \rangle$ is called a \textit{subsequence} of $s$ and $s$ is called a \textit{supersequence} of $p$, denoted as $p \sqsubseteq s$, if there exists a sequence of consecutive integers $1 \leq B_1 < B_2 < \ldots < B_l \leq h$ such that $A_1=L_{B_1}, A_2=L_{B_2}, \ldots, A_l=L_{B_l}$. A route $\mathcal{T}$ is said to \textit{contain} $p$, if $s$ is induced by $\mathcal{T}$ and $p \sqsubseteq s$, which is denoted as $p \prec \mathcal{T}$ for simplicity. $p$ is said to be a \textit{$l$-pattern}, if $len(p)=l$.

The \textit{support of a sequence} $p$ in $\mathcal{D}$ is the number of routes containing $p$, i.e., $f_\mathcal{D}(p)=|\{\mathcal{T}|p \prec \mathcal{T} \wedge \mathcal{T} \in \mathcal{D}\}|$. Given a positive integer $min\_sup$ as the \textit{sequence support threhold}, a sequence $p$ is called a \textit{sequential pattern} in $\mathcal{D}$ if $f_\mathcal{D}(p) \geq min\_sup$. 

Now we give an example to explain the concepts, including the graph, the route, the sequence and the pattern.

\begin{example}[Concepts]
\label{ex:concepts}
Figure~\ref{fig:ex} and Figure~\ref{fig:route} show a tiny graph and a set of routes. The graph $G$ has 10 vertices $\{v_1, \ldots, v_{10}\}$, 17 edges and $\Pi=\{PS, WS, MS, DB\}$. The route set $\mathcal{D}=\{\mathcal{T}_1, \ldots, \mathcal{T}_8\}$. We can always find a sequence supporting a route. For example, $s_1=\langle PS, WS, DB \rangle$ is induced by $\mathcal{T}_1$. Given $min\_sup=3$, there are 4 frequent sequential patterns, whose lengths are greater than or equal to 2. The patterns are $\langle PS, MS \rangle$, $\langle PS, MS, DB \rangle$, $\langle PS, DB \rangle$ and $\langle MS, DB \rangle$. Suppose that $H_2$ is a subgraph induced by $\{v_4, v_6, v_7, v_8, v_9, v_{10}\}$, and it's obvious that $H_2$ is a $3$-truss and is \textit{maximal}.
\end{example}

\subsection{Route Hotspot}

With the commonly accepted desiderata of a good route hotspot described in Section~\ref{sec:introduction}, we now give a precise definition of \textit{route hotspot}.

\begin{definition}[Route Hotspot]
\label{def:communityDef}
Given a graph $G$, a set of routes $\mathcal{D}$, a pattern $p$, two parameters $k$ and $min\_sup$, $H_p(k)$ (denoted as $H$ for short) is a route hotspot, if $H$ satisfies the following conditions:
\begin{enumerate}

	\item $H$ is a $k$-truss and is connected.
	
	\item $H$ contains at least $min\_sup$ subroutes of routes in $\mathcal{D}$, where each subroute $t$ contains the same pattern $p$. Formally, $\exists \mathcal{D}_p=\{t \in \mathcal{T}|p \prec t \wedge \mathcal{T} \in \mathcal{D}\}$ and $f_{\mathcal{D}_p}(p) \geq min\_sup$, such that $\mathcal{D}_p \Subset H$.
	
	\item $\mathcal{D}_p$ can cover $H$ and the length of each subroute $t$ in $\mathcal{D}_p$ should be as long as possible. Formally, $H \boxdot \mathcal{D}_p$.
    
         \item $H$ is a maximal subgraph satisfying conditions (1), (2) and (3). That is, $\nexists H' \subseteq G$, such that $H \subset H'$, and $H'$ satisfies conditions (1), (2) and (3). 

\end{enumerate}
\end{definition}

Note that we don't consider the patterns whose lengths are equal to 1 since it's trivial and meaningless. An example of route hotspot $H$ for $p=\langle PS, MS \rangle$ and $k=3$ is given in Section~\ref{sec:introduction}.1, with the corresponding $G$ in Figure~\ref{fig:ex}, $\mathcal{D}$ in Figure~\ref{fig:route} and $min\_sup=3$ respectively.

Note that a vertex whose label doesn't belong to the pattern may also appear in a hotspot, in order to act as a ``bridge" between other vertices. Specifically, suppose $p \prec \mathcal{T}$, the length of $\mathcal{T}$ may be longer than the length of $p$ and thus some vertices in $\mathcal{T}$ may be associated with labels that don't exist in $p$. For example, in Figure~\ref{fig:route}, $\mathcal{T}_6$ contains the pattern $\langle PS, MS \rangle$, but $v_9 \in \mathcal{T}_6$ doesn't correspond to the label $PS$ or $MS$. Nevertheless, $v_9$ is still involved in the hotspot as it may be indispensable in a practical network, e.g., it may act as a starting point or an endpoint of an attack sequence in the network security scenario. Moreover, longer routes are expected to cover route hotspots more easily than those with shorter lengths, and thus can help us find more hotspots.

\subsection{The FRHLN problem}

Now we are ready to define the problem of \textit{Finding Route Hotspots in Large Labeled Networks (FRHLN) }. 

\begin{definition}[FRHLN]
\label{def:problemDef}
Given a graph $G$, a route set $\mathcal{D}$, a user-specific parameter $min\_sup$, the problem of  \textit{finding route hotspots in large labeled networks} is to enumerate all route hotspots $H_p(k)$ for each $p$ and $k$, where $k \geq 2$, $len(p) \geq 2$ and each $H_p(k)$ contains at least $min\_sup$ subroutes.
\end{definition}

We prove in Theorem~\ref{th:hardness} that counting the number of the route hotspots is \#P-hard.

\begin{theorem}[Hardness]
\label{th:hardness}
Given a graph $G$, a route set $\mathcal{D}$ and a user input threshold $min\_sup$, the problem of counting the number of route hotspots in $G$ is \#P-hard.
\end{theorem}

\begin{IEEEproof}[Sketch]
We prove this theorem by a reduction from the \textit{Sequential Pattern Mining} (SPM) problem, which is known to be \#P-hard~\cite{Gunopulos2003}. Given an instance of the SPM problem, where the input is a sequence database $S$ and a minimal support $min\_sup$, we are asked how many patterns $p$ in $S$ satisfy $f_{S}(p) \geq min\_sup$. Here,  $f_{S}(p)$ is the number of unique sequences in $S$ that contain $p$. We denote the union of the items (i.e., the labels of the vertices) in $S$ as $\mathbb{I}$, where the items follow an arbitrary fixed total order. Using $S$, we can construct a graph $G$ and a route set $\mathcal{D}$ as follows. 

We assume that each item in $\mathbb{I}$ represents a vertex (say $v_i$) in $G$ and with a unique label (say $\mathbb{I}_i$). Therefore, the number of vertices and labels in $G$ are both $|\mathbb{I}|$. To construct $G$, we add edges for every two vertices, and then every three vertices can form a triangle. For each sequence $s$ in $S$, we build a route $\mathcal{T}$ with the following two steps. \text{Step 1:} since $s$ is an ordered list of items, let $\mathcal{T}$ travel the edges with the order of items in $s$. For example, if $s=\langle \mathbb{I}_4, \mathbb{I}_2, \mathbb{I}_3 \rangle$, $\mathcal{T}$ travels $(v_4, v_2)$ and $(v_2, v_3)$ sequentially. \text{Step 2:} let $\mathcal{T}$ travel all the other edges which are not visited in Step 1. Apparently, $G$ is ensured to be covered by $\mathcal{T}$. Then we obtain a route for each $s$, and put all these routes in $\mathcal{D}$. This reduction step takes polynomial time.

For any pattern $p \subset S$, since $f_{S}(p) \geq min\_sup$, at least $min\_sup$ routes in $\mathcal{D}$ contain $p$. As $G$ is a complete graph, according to our route hotspot definition, there must exist a route hotspot $H_p(k)$ for a pattern $p$ and a $k$, where $k$ ranges from 2 to $|\mathbb{I}|-1$. Therefore, the number of route hotspots in $G$ is $|\mathbb{I}|-2$ times the number of patterns in $S$, which is exactly the answer to the SPM problem. Consequentially, as SPM problem, the problem of counting the number of route hotspots in $G$ is also \#P-hard.
\end{IEEEproof}

The above theorem shows that finding a polynomial-time exact algorithm for computing the number of route hotspots in $G$ is hard. Thus we need to look for pruning strategies to tackle the efficiency issue.

Table~\ref{tab:notations} summarizes the frequently used notations for the rest of the paper.

\section{The Greedy Route Hotspot Finding}
\label{sec:simpleAl}
In this section, we first define the trussness of the subgraph, edge, and route, and then introduce the greedy route hotspot detection method. 

\subsection{The Trussness of Subgraph, Edge and Route}
The definitions of the trussness of subgraph, edge, and route are listed as follows.

\begin{definition}[Subgraph Trussness~\cite{Huang2014}]
\label{def:subgraphTrussness}
The trussness of a subgraph $H \subseteq G$ represents the minimum support of an edge in $H$, denoted by $\tau(H)$. For the purpose of unifying the notion of $k$-truss, we make $\tau(H)=min\{sup(e, H)| e \in E(H)\}+2$. Thus, the trussness of a $k$-truss is equal to $k$. 
\end{definition}

\begin{definition}[Edge Trussness~\cite{Huang2014}]
\label{def:edgeTrussness}
The trussness of an edge $e \in E(H)$ is defined as $\tau_H(e)=max_{H' \subseteq H}\{\tau(H') | e \in E(H')\}$.
\end{definition}

\begin{definition}[Route Trussness]
\label{def:routeTrussness}
The trussness of a route $\mathcal{T} \in H$ is the maximum trussness of the edge trussness, denoted by $\tau_H(\mathcal{T})=max\{\tau_H(e)|e \in \mathcal{T}\}$.
\end{definition}

Consider the graph $G$ in Figure~\ref{fig:ex} as an example. The trussness of $G$ is $\tau(G)=2$. According to Definition~\ref{def:edgeTrussness}, the trussness of an edge $e$ is equal to the maximum value of the trussness for different subgraphs that contain $e$. For example, the subgraph $H=(V_H, E_H)$ contains the edge $e(v_4, v_9)$, where $V_H=\{v_4, v_6, v_7, v_8, v_9, v_{10}\}$ and $E_H=\{e(v_4, v_6), e(v_4, v_7), e(v_4, v_9), e(v_6, v_7), e(v_6, v_8),$ $e(v_7, v_8), e(v_7, v_9), e(v_7, v_{10}), e(v_8, v_{10})\}$. The trussness of $H$ is 4, which is the largest value among that of subgraphs containing $e(v_4, v_9)$. Thus, the trussness of the edges $e(v_4, v_9)$ is 4. The trussness of the routes $\mathcal{T}_1$, $\mathcal{T}_5$ in Figure~\ref{fig:route} are $\tau_G(\mathcal{T}_1)=2$, $\tau_G(\mathcal{T}_5)=4$. 

The above definitions can be leveraged to prune unqualified edges and routes. We say that, an edge or a route is \textit{unqualified} if it cannot be used to form a route hotspot. We also say that, a pair of pattern $p$ and $k$ is \textit{unqualified} if $H_p(k)=\emptyset$. For instance, for a route hotspot $H_p(k)$, the support of each edge in $H_p(k)$ must be greater than or equal to $k$. Thus, if $\tau_G(\mathcal{T})<k$, which indicates that the trussness of edges in $\mathcal{T}$ is less than $k$, $\mathcal{T}$ is thus unqualified for a route hotspot and can be pruned safely.

\begin{algorithm}[t]
\algsetup{linenosize=\small}
\scriptsize
\caption{\small{Hotspot Detection For $p$ And $k$}}
\label{al:singleAl}
\KwIn{$H_p(V_p, E_p)$, $\mathcal{D}_p$, $min\_sup$, $p$, $k$}
\KwOut{A set of hotspots $\Phi_p(k)$}
\begin{algorithmic}[1]
\STATE $Q \gets \emptyset$;
\STATE Compute $sup(e, H_p)$ for each edge $e \in E_p$; 
\STATE Push each $e(u,v)$ into $Q$ if $sup(e, H_p)<(k-2)$;
\WHILE{$Q \neq \emptyset$}
	\STATE $e(u,v)=Q.dequeue()$;	
	\FOR{Each $w \in N(u)$ and $(v, w) \in E_p$}
		\STATE $sup((u,w), H_p) \gets sup((u,w), H_p)-1$;
		\STATE $sup((v,w), H_p) \gets sup((v,w), H_p)-1$;
	\ENDFOR
	\WHILE{$\exists \mathcal{T} \in \mathcal{D}_p$, such that $e \in \mathcal{T}$} 
		\STATE Remove $e$ from $\mathcal{T}$;
	\ENDWHILE
	\WHILE {$\exists \mathcal{T} \in \mathcal{D}_p$, such that $\tau_{H_p}(\mathcal{T}) < k \vee p \notin \mathcal{T}$}
		\STATE Remove $\mathcal{T}$ from $\mathcal{D}_p$;
	\ENDWHILE
	\STATE Remove $e$ from $E_p$;
	\WHILE{{$\exists e' \in E_p$, such that $e' \notin$ any $\mathcal{T}$ in $D_p$}}
		\STATE $sup(e', H_p)=0$;		
	\ENDWHILE
	\STATE Push each $e'$ into $Q$ if $sup(e', H_p)<(k-2)$;
\ENDWHILE
\IF{$E_p \neq \emptyset$}
	\WHILE{$\exists$ a connected subgraph $H_p(k)$ of $H_p$, such that $\mathcal{D}_p(k) \Subset H_p(k)$ and $|\mathcal{D}_p(k)| \geq min\_sup$}			
		\STATE $\Phi_p(k) \gets \Phi_p(k) \cup H_p(k)$;
	\ENDWHILE		
\ENDIF
\RETURN $\Phi_p(k)$
\end{algorithmic}
\end{algorithm}

We then propose Algorithm~\ref{al:singleAl} to detect hotspots for a pair of inputs $p$ and $k$. The general idea is that, we iteratively prune unqualified edges and routes based on the trussness information, and use the remaining edges to form the route hotspots. Specifically, let $H_p(V_p, E_p)$ denote a subgraph induced by $\mathcal{D}_p$. We first set up an empty queue $Q$ to store unqualified edges (Line 1). We compute the support for each edge and put unqualified edges, whose supports are less than $k-2$, into $Q$ (Lines 2-3). For each $e$ in $Q$, we decrement the supports of those related edges who can form triangles with $e$ in $H_p$ (Lines 6-9). Then $e$ will be removed from the routes that contain $e$ (Lines 10-12). Upon the removal of $e$, some routes may become unqualified and thus will be removed from $\mathcal{D}_p$ (Lines 13-15). After that, $e$ is removed from $E_p$ (Line 16). For each $e' \in E_p$, if it is not contained in any $\mathcal{T}$, the support of $e'$ will be set as zero (Lines 17-19). We then find all the edges that are unqualified again and put them into $Q$ (Line 20). This process iterates until $Q$ becomes empty. The remaining edges in $E_p$ are used to form connected subgraphs. If the subgraph contains at least $min\_sup$ routes that contain $p$, it would be a route hotspot and put into $\Phi_p(k)$ (Lines 22-26). Finally, $\Phi_p(k)$ is returned (Line 27). 

The correctness of Algorithm~\ref{al:singleAl} is apparent since the algorithm essentially detects hotspots by definition. The time complexity of Algorithm~\ref{al:singleAl} is determined by the enumeration of all triangles for each edge in $H_p (V_p, E_p)$. The time complexity of this enumeration is $\mathcal{O}(\sum_{(v_i, v_j) \in E_p}(d(v_i)+d(v_j)))=\mathcal{O}(|V_p| d_{pmax}^2)$, where $|V_p|$ is the volume of $V_p$ and $d_{pmax}$denotes the maximum degree of the vertices in $V_p$.

\subsection{The Greedy Route Hotspot Detection (GreedyRH)}
Algorithm~\ref{al:singleAl} can be used iteratively to find all route hotspots, but it requires all feasible patterns and feasible values of $k$ as prior inputs. To obtain such inputs, we denote $S=\{s \prec \mathcal{T}|\mathcal{T} \in \mathcal{D}\}$ as a set of sequences, and denote $P$ as a set of patterns mined from $S$ with the support threshold $min\_sup$. We also denote $k_{max}-truss$ as the \textit{maximum truss} in $G$, indicating that $\nexists H \in G$, such that $\forall e \in E_{H}, sup(e, H) \geq (k_{max}-1)$. 

\begin{observation}
\label{ob:patternFrequency}
If there exists a route hotspot $H_p(k)$, $p$ must belong to $P$.
\end{observation}

\begin{observation}
\label{ob:k}
For each route hotspot $H_p(k)$, $k$ must be less than or equal to $k_{max}$.
\end{observation}

Observation~\ref{ob:patternFrequency} implies that $P$ is a superset of qualified patterns. Observation~\ref{ob:k} implies that $k_{max}$ is the upper bound of the maximum value of $k$. Based on these observations, we propose Algorithm~\ref{al:simpleAl}, called \textit{GreedyRH}, that runs iteratively for each pattern $p$ and each $k$ ($2 \leq k \leq k_{max}$) and returns all route hotspots. $P$ can be obtained by applying a sequential pattern mining method, like PrefixSpan~\cite{Pei2001}. And $k_{max}$ can be obtained by applying a truss decomposition method, like~\cite{Wang2012}.

\begin{algorithm}[t]
\algsetup{linenosize=\small}
\scriptsize
\caption{\small{Greedy Hotspot Detection (GreedyRH)}}
\label{al:simpleAl}
\KwIn{Graph $G$, $\mathcal{D}$, $min\_sup$, $P$, $k_{max}$}
\KwOut{A set of hotspots $\Phi$}
\begin{algorithmic}[1]
\FOR{Each $p \in P$}
	\FOR{$k=2 \to k_{max}$}
		\STATE Initialize $\mathcal{D}_p$ based on $\mathcal{D}$ and induce $H_p(V_p, E_p)$ based on $\mathcal{D}_p$;
		\STATE $\Phi_p(k)$:= Call Algorithm~\ref{al:singleAl} by using $H_p$, $\mathcal{D}_p$, $min\_sup$, $p$, $k$ as input;
		\STATE $\Phi \gets \Phi \cup \Phi_p(k)$;	
	\ENDFOR
\ENDFOR
\RETURN $\Phi$
\end{algorithmic}
\end{algorithm}

The time complexity of GreedyRH is related to the number of patterns and $k_{max}$. We denote the maximal length of routes as $h_{max}$. Thus, the number of patterns is at most $|\Pi|^{h_{max}}$. For each pattern, the Algorithm~\ref{al:singleAl} will be performed for at most $(k_{max}-1)$ times. Therefore, the time complexity of GreedyRH is $\mathcal{O}(|\Pi|^{h_{max}} \times(k_{max}-1)\times|V| \times d_{max}^2)$, where  $|V|$ is the volume of $V$ and $d_{max}$ denotes the maximum degree of the vertices in $V$. 

The greedy algorithm, i.e., GreedyRH, is correct due to the following reasons. First, it calls Algorithm~\ref{al:singleAl} iteratively to find all route hotspots, where Algorithm~\ref{al:singleAl} detects a specific route hotspot for a pair of inputs $p$ and $k$ by definition. Second, according to Observations~\ref{ob:patternFrequency} and~\ref{ob:k}, we can obtain all the patterns and the maximum value of $k$. Thus, GreedyRH can find all qualified route hotspots for all the patterns and all possible values of $k$, ensuring the correctness of this method. However, running GreedyRH is impractical. We then analyze the unnecessary computational overhead.

\begin{enumerate}
\item \textbf{Not every $k$-truss can form a hotspot}: The upper bound $k_{max}$ for $k$ may be too loose. When increasing $k$, some edges may be removed from $G$ and the remaining part may not be covered by at least $min\_sup$ routes. Thus, checking every $k$ \textit{becomes unnecessary}.

\item \textbf{Not every pattern is frequent in a $k$-truss}: In a $k$-truss of $G$, the number of routes containing the same $p$ may be less than $min\_sup$. Thus, checking unfrequent patterns in $k$-trusses also \textit{becomes unnecessary}.
\end{enumerate}

To overcome these drawbacks, we develop an improved algorithm and describe it precisely in the next section.

\section{The Novel Route Hotspot Finding}
\label{sec:novelAl}
In this section, we first prove the anti-monotonicity and independence properties that can be used to avoid unnecessary checking. With these 
properties, we then propose a novel efficient finding method.

\subsection{Anti-Monotonicity And Independence Properties}

\begin{theorem}[Pattern Anti-monotonicity on Hotspots]
\label{th:pattern}
For patterns $p_1$, $p_2$, where $|p_1| \geq 2$ and $p_1 \sqsubseteq p_2$, and a fixed $k$, where $k \geq 2$, the following two properties hold. 
\begin{enumerate}
\item If there exists a hotspot $H_{p_2}(k) \neq \emptyset$, there must exist a hotspot $H_{p_1}(k) \neq \emptyset$.

\item If  $H_{p_1}(k) = \emptyset$, $H_{p_2}(k)= \emptyset$.
\end{enumerate}
\end{theorem}

\begin{IEEEproof}
Suppose that $H_{p_2}(k)=(V_{p_2}(k), E_{p_2}(k))$ is a hotspot for $p_2$. We build a subgraph $H_{p_1}=(V_{p_1}, E_{p_1})$, where $V_{p_1}=V_{p_2}(k)$ and $E_{p_1}=E_{p_2}(k)$. There must exist $min\_sup$ routes, denoted by $\mathcal{D}'$, containing $p_2$. Since $p_1 \sqsubseteq p_2$, for $\forall \mathcal{T} \in \mathcal{D}'$, $p_1 \prec \mathcal{T}$. Thus, $\mathcal{D}'$ can also cover $H_{p_1}$. For $p_1$ and $k$, $H_{p_1}$ and $\mathcal{D}'$ satisfy the conditions (1), (2) and (3) of Definition~\ref{def:communityDef}, respectively. We just need to determine whether there exists a subgraph $H_{p_1}(k)$ ($H_{p_1} \subset H_{p_1}(k)$) satisfying condition (4) of Definition~\ref{def:communityDef}. As $p_1 \sqsubseteq p_2$, the size of $H_{p_1}$ must be larger than or equal to the size of $H_{p_2}(k)$, i.e., $H_{p_2}(k) \subseteq H_{p_1}$. Meanwhile, it is obvious that $H_{p_1} \subseteq G$. Thus, we can always find a maximal $H_{p_1}$ (say $H_{p_1}(k)$) which is exactly a route hotspot for $p_1$ and $k$, where $H_{p_1} \subseteq H_{p_1}(k)$ and $H_{p_1}(k) \subseteq G$. The property 1 holds.

The second property is the contraposition of the first one, and thus it holds as well.
\end{IEEEproof}

\begin{theorem}[Hotspot Anti-monotonicity with Patterns]
\label{th:graph}
For a pattern $p$ and a fixed $k$ $(k>2)$, if there exists a hotspot $H_p(k)$, there must exist a hotspot $H_p(k-1)$.
\end{theorem}

\begin{IEEEproof}
Suppose that $H_p(k)=(V_p(k), E_p(k))$ is a hotspot for $p$ and $k$. We build a subgraph $H_p=(V_p, E_p)$, where $V_p=V_p(k)$ and $E_p=E_p(k)$. $H_p$ is a $k$-truss, where $sup(e,H) \geq (k-1) \geq (k-2)$. Thus, $H_p$ is also a $(k-1)$-truss.  Since $H_p(k)$ is a route hotspot, there must exist $min\_sup$ routes, denoted by $\mathcal{D}'$, containing $p$. $\mathcal{D}'$ can also cover $H_p$. For $p$ and $k-1$, $H_p$ and $\mathcal{D}'$ satisfy the conditions (1), (2) and (3) of Definition~\ref{def:communityDef}, respectively. We just need to determine whether $H_p$ satisfies condition (4) of Definition~\ref{def:communityDef}, i.e., whether $H_p$ is a maximal subgraph for $p$ and $(k-1)$. However, the size of $H_p$ is at least equal to the size of $H_{p}(k)$. Similar to the proof for Theorem~\ref{th:pattern}, there must exist a $H_p$ satisfying condition (4) of Definition~\ref{def:communityDef}, which is exactly a hotspot $H_{p}(k-1)$ for $p$ and $(k-1)$.
\end{IEEEproof}

The above two properties ensure that we can prune unqualified patterns and values of $k$ efficiently and safely. Theorem~\ref{th:pattern} indicates how to prune unqualified edges for a new pattern based on existing patterns. For any pattern $p_1$ and $p_2$, if $p_1 \sqsubseteq p_2$ and $p_1$ is unqualified, then $p_2$ is also unqualified. Theorem~\ref{th:graph} indicates how to prune unqualified edges for the same pattern with different values of $k$. For any $p$, if $k$ is unqualified, then $(k+1)$ is also unqualified.

We then give an example to show how the above two theorems work. Continued with Example~\ref{ex:concepts}, we consider $min\_sup=3$. According to Theorem~\ref{th:pattern}, $\langle WS \rangle$ cannot be grown to $\langle WS, DB \rangle$, since there are only 2 routes containing $\langle WS, DB \rangle$, which doesn't meet the condition of $min\_sup$. Then, there are no route hotspots for $\langle WS, DB \rangle$ or other patterns grown from $\langle WS, DB \rangle$. According to Theorem~\ref{th:graph}, given the pattern $\langle PS, MS \rangle$, as there are no route hotspots for $k=4$, there will be no route hotspots for $k=5$ or even larger values of $k$.

We also propose the independence property for route hotspot detection. 

\begin{proposition}[Computation Independence]
\label{prop:distributed}
The detection process for different patterns is independent.
\end{proposition}

Proposition~\ref{prop:distributed} allows us to speed up the route hotspot detection in a parallel manner. Next, based on Theorem~\ref{th:pattern} and~\ref{th:graph} as well as Proposition~\ref{prop:distributed}, we will introduce the improved hotspot finding algorithm compared with GreedyRH.

\begin{algorithm}[t]
\algsetup{linenosize=\small}
\scriptsize
\caption{\small{Fast Hotspot Scanner (FastRH)}}
\label{al:thd}
\KwIn{$G$, $\mathcal{D}$, $min\_sup$}
\KwOut{The set of hotspots $\Phi$}
\begin{algorithmic}[1]
\STATE Initialize: $P^2$, $Q_p \gets \emptyset$;
\STATE Put all $p^2 \in P^2$ into $Q_p$;
\WHILE{$Q_p \neq \emptyset$}
	\STATE $p^i=Q_p.dequeue()$, $k \gets 2$;
	\STATE Initialize $\mathcal{D}_{p^i}$ based on $\mathcal{D}$ and induce $H_{p^i}(V_{p^i}, E_{p^i})$ based on $\mathcal{D}_{p^i}$;
	\WHILE{$E_{p^i} \neq \emptyset \wedge |\mathcal{D}_{p^i}| \geq min\_sup$}
		\STATE $\Phi_{p^i}(k)$:= Call Algorithm~\ref{al:singleAl} by using $H_{p^i}$, $\mathcal{D}_{p^i}$, $min\_sup$, ${p^i}$, $k$ as input;
		\STATE $\Phi \gets \Phi \cup \Phi_{p^i}(k)$;			
		\STATE $k \gets k+1$;
	\ENDWHILE
	\IF{$\Phi_{p^i}(2) \neq \emptyset$}
		\STATE Generate $P^{i+1}$ based on $p^i$;
		\STATE Put all $p^{i+1} \in P^{i+1}$ into $Q_p$;
	\ENDIF	
\ENDWHILE
\RETURN $\Phi$
\end{algorithmic}
\end{algorithm}

\subsection{The Fast Route Hotspot Scanner (FastRH)}

Algorithm~\ref{al:thd} shows the procedure of the improved route hotspot detection, i.e., \textit{FastRH}, that can prune unnecessary checking for unqualified patterns and values of $k$. Denote $p^i$ as the length-$i$ pattern. For initialization, we first calculate the set of length-2 patterns $P^2$ (Line 1) and put them into a pattern queue $Q_p$ (Line 2). For each pattern $p^i$ extracted from $Q_p$ (Line 4), we initialize the route set $\mathcal{D}_{p^i}$ for $p^i$ and build a graph $H_{p^i}$ based on $\mathcal{D}_{p^i}$ (Line 5).  We iteratively call Algorithm~\ref{al:singleAl} to obtain the route set $\Phi_{p^i}(k)$. According to Theorem~\ref{th:graph}, instead of enumerating $k$ from 2 to $k_{max}$ in Algorithm~\ref{al:simpleAl}, here we increment $k$ until all the edges have been removed (Lines 6-10), which can reduce the number of iterations. According to Theorem~\ref{th:pattern}, instead of enumerating patterns from $P$ in Algorithm~\ref{al:simpleAl}, here we generate $P^{i+1}$ only when the hotspots for $p^i$ exist, and put all candidate length-$(i+1)$ patterns into $Q_p$; otherwise, we stop the pattern growth from $p^i$ immediately (Lines 11-14). This process iterates until $Q_p$ becomes empty. Finally, the exact set of route hotspots $\Phi$ is returned. Please note that lines 3-15 can be parallelized according to Proposition~\ref{prop:distributed}. That is, we use the multithreading technique and process multiple patterns at a time in each iteration.

Compared with GreedyRH, the scanner is improved in efficiency due to two reasons. First, it can reduce unnecessary checking for unqualified patterns since a large number of patterns that cannot form route hotspots have been pruned (Line 11-14). Theoretically, we can have up to $|\Pi|^{h_{max}}$ patterns, where $h_{max}$ is the maximum length of patterns. E.g., the datasets GW in Section~\ref{sec:experiments} has $10^7$ patterns, but only $8.6 \times 10^5$ patterns (in Table~\ref{tab:indexStatistics}) can form route hotspots. That means FastRH prunes 91.4\% patterns safely. Second, FastRH saves unnecessary checking for unqualified values of $k$. For instance, Table~\ref{tab:indexStatistics} demonstrates that the maximum $k$ for GW is 6, less than $k_{max}=16$ in Table~\ref{tab:graphStatistics}, indicating the unnecessary computation with $k$ ranging from 7 to 16. Please note that FastRH and its parallelized version PFastRH can produce the correct results of route hotspot finding, as they both adopt Theorems~\ref{th:pattern} and~\ref{th:graph} that can prune unqualified patterns and values of $k$ safely.

Although the computation cost can be saved significantly with FastRH, when a user inputs a new $min\_sup$, the route hotspot detection procedures have to be performed from scratch, even if the same detection has been performed before. To avoid the duplicate detection, it is essential to store the detected hotspots to enable the quick answering for new user queries. In the next section, we develop a novel framework to index all the detected route hotspots.

\section{Route Hotspot Indexing}
\label{sec:index}
In this section, we propose an efficient indexing structure for the route hotspots, which is named as the \textit{Route Hotspot Index}, or RH-Index for short.

The \textit{RH-Index} follows the pattern growth property, which is first introduced in FreeSpan~\cite{Pei2000}. However, the existing pattern growth-based tree structures like FreeSpan can only mine and index the sequential patterns whose information is quite simple, but fall short in storing the rich information associated with route hotspots. In particular, they are not able to store the hotspot information and thus cannot be applied to recover the route hotspots from the index for answering the user queries. Moreover, their index construction processes only consider the pattern anti-monotonicity, but ignore the hotspot anti-monotonicity, which may result in low efficiency due to the redundant computing for unqualified values of $k$. 

To overcome these challenges, RH-Index is proposed by introducing the hotspot information and qualified values of $k$ in the index structure. Compared to previous pattern growth-based tree structures such as FreeSpan, the benefits are two-fold. First, it can recover the indexed route hotspots with the linked edges stored in the index. Moreover, the recovery procedure can be quite efficient by using the stored trussness information of edges. Specifically, the RH-Index can directly use the edges, whose trussness is greater than or equal to $k$, to recover the indexed route hotspots, avoiding extracting unqualified edges whose trussness is smaller than $k$. Second, following the pattern and hotspot anti-monotonicity properties, the RH-Index can update incrementally when new routes are inserted or the existing routes are deleted, minimizing the overhead needed for index reconstruction.

\subsection{Route Hotspot Index Construction}
\begin{figure}[t]
\centering
	\includegraphics[scale=0.3]{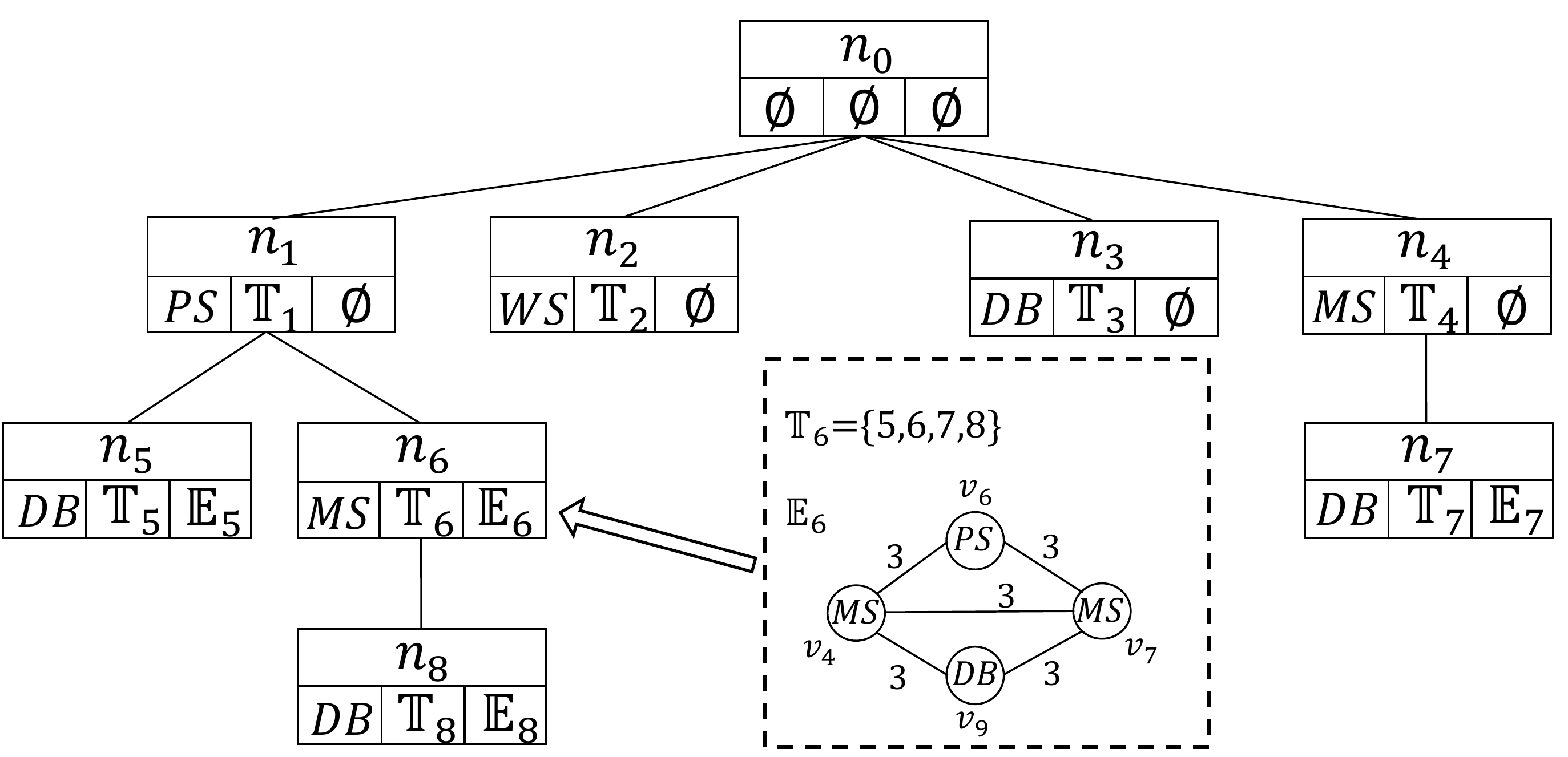}
\caption{An example of RH-Index. Note that, although the first layer nodes, such as $n_2$ and $n_4$, cannot form any route hotspots, we still retain these nodes to support index updating.}
\label{fig:thIndex}
\end{figure}

We first define RH-Index. Denote $\mathbb{L}$ as the set of all items in the patterns, and we map each item in $\mathbb{L}$ to $\Pi$ by using a function $\chi: \mathbb{L} \rightarrow \Pi$. The $i$-th node of the RH-Index is denoted by $n_i$, consisting of the item $\mathbb{L}_i$, the list of routes $\mathbb{T}_i$ and the list of edges $\mathbb{E}_i$. The $i$-th pattern $p_i$ is the inverted sequence traveling from $n_i$ to $n_0$ along the tree structure. $\mathbb{T}_i$ stores the routes which can cover all the hotspots for $p_i$. The edges of the found hotspots are stored in $\mathbb{E}_i$. The edges in $\mathbb{E}_i$ are the tuples, each of which consists of an edge $e$ and the edge weight $w(e)$, where $w(e)$ represents the maximal trussness of the hotspot containing $e$. Given $G$, $\mathcal{D}$ in Example~\ref{ex:concepts} and $min\_sup=3$, we can build the corresponding RH-Index consisting of 8 nodes, as shown in Figure~\ref{fig:thIndex}. For node $n_0$, $\mathbb{L}_0=\emptyset$, $\mathbb{T}_0=\emptyset$, $\mathbb{E}_0=\emptyset$. For the $1$-pattern $p$,  $\mathbb{T}_{p}=\emptyset$, $\mathbb{E}_{p}=\emptyset$, since we only consider the patterns whose lengths are greater than 1. $n_6$ corresponds to indexed hotspots for $p_6=\langle PS, MS \rangle$. $\mathbb{L}_6=MS$, $\mathbb{T}_6=\{5,6,7,8\}$, $\mathbb{E}_6=\{((v_4, v_6),3), ((v_4, v_7),3), ((v_4, v_9),3), ((v_6, v_7), 3), ((v_7, v_9), \\ 3)\}$. 

\begin{algorithm}[t]
\algsetup{linenosize=\small}
\scriptsize
\caption{\small{RH-Index Construction}}
\label{al:index}
\KwIn{$G$, $\mathcal{D}$, The set of route hotspots $\Phi$}
\KwOut{A RH-Index $\mathbb{T}$}
\begin{algorithmic}[1]
\STATE $Q \gets \emptyset$, $n_0.\mathbb{L}_0 \gets \emptyset$, $n_0.\mathbb{E}_0 \gets \emptyset$ and $n_0.\mathbb{T}_0 \gets \emptyset$;
\FOR{Each item $L_i \in L$}
	\STATE Assign $n_i.\mathbb{L}_i \gets L_i$, $n_i.\mathbb{E}_i \gets \emptyset$ and calculate $n_i.\mathbb{T}_i$;
	\STATE $n_i.p_i \gets L_i$;
	\STATE $n_0.addChild(n_i)$, $Q.inqueue(n_i)$;
\ENDFOR
\WHILE{$Q \neq \emptyset$}
	\STATE $n_i=Q.dequeue()$;
	\IF{$n_i.\mathbb{E}_i \neq \emptyset$}
		\FOR {Each item $L_j \in L$}
			\STATE $p \gets n_i.p_i \cup L_j$;
			\STATE $\mathbb{L}_j=L_j$;		
			\STATE $\mathbb{T}_j \gets \mathcal{D}_{H_p(2)}$, where $H_p(2) \in \Phi_p(2)$;
			\STATE $\mathbb{E}_j \gets  \{(e, k)|e \in E_{H_p(k)} \wedge H_p(k) \in \Phi_p(k)\}$;
			\STATE $n_i.addChild(n_j)$;$Q.inqueue(n_j)$;
		\ENDFOR	
	\ENDIF
\ENDWHILE
\RETURN $\mathbb{T}$
\end{algorithmic}
\end{algorithm}

We then introduce how to construct the RH-Index with the given $G$, $\mathcal{D}$ and the returned route hotspots $\Phi$. The index is built in a top-down manner by using Algorithm~\ref{al:index}. We first initialize a queue $Q$ and construct the root node of $\mathbb{T}$, where $\mathbb{L}_0=\emptyset$, $\mathbb{T}_0=\emptyset$, $\mathbb{E}_0=\emptyset$ (Line 1). Then, we generate the first layer nodes of $\mathbb{T}$, which represent length-$1$ patterns, and put all nodes into $Q$ (Lines 2-6). Here $n_i.\mathbb{L}_i$ and $n_i.\mathbb{T}_i$ refer to $\mathbb{L}_i$ and $\mathbb{T}_i$ of node $n_i$, respectively. The rest of RH-Index nodes are built iteratively as the lengths of patterns increase. For each node $n_i$ dequeued from $Q$, we first obtain the pattern $p$ represented by $n_i$ and then grow $p$ by appending $L_j$ (Line 11). Then $\mathbb{L}_j$, $\mathbb{T}_j$ and $\mathbb{E}_j$ of $n_j$ will be built correspondingly (Lines 12-15). Finally, the RH-Index $\mathbb{T}$ is successfully built and returned. Please note that the construction of RH-Index don't need to wait until FastRH has detected all the route hotspots. Instead, they may interleave to improve the efficiency. In particular, each time when the set of route hotspots for a specific pattern $p$ is found, a new tree node $n_p$ can thus be built on the tree.

When new routes are inserted or existing routes are deleted, the RH-Index is required to be updated. Inserting new routes may lead to (1) updating some existing nodes with the revised trussness information and edges; or (2) inserting some new nodes as some new patterns become qualified. Let $\mathcal{D}_{new}$ denote the set of inserted routes, and Algorithm~\ref{al:singleAl} is called to determine the nodes that represent the patterns contained by $\mathcal{D}_{new}$ and make the corresponding updating operations. Note that the pattern and hotspot anti-monotonicity properties introduced in Theorem~\ref{th:pattern} and Theorem~\ref{th:graph} can be used to prune unqualified patterns and values of $k$, avoiding unnecessary updates. This updating process continues until pattern growth ends. We omit the route deletion process as it is similar to the insertion case.

\subsection{Query Processing Using RH-Index}
\label{sec:updating}

We can easily retrieve the hotspots with the given $p$ and $k$, since the nodes of RH-Index store the trussness information as the weights of edges, which indicate the largest values of $k$ for each hotspot marked by $p$. Here we propose a query processing method for RH-Index. The \textit{answer} to query $(p, k)$ is the set corresponding to hotspots $\Phi_p(k)$.

\begin{algorithm}[t]
\algsetup{linenosize=\small}
\scriptsize
\caption{\small{Query Processing with RH-Index}}
\label{al:query}
\KwIn{$\mathbb{T}$, $p$, $k$}
\KwOut{The set of hotspots $\Phi_p(k)$}
\begin{algorithmic}[1]
\STATE $Q \gets n_0$, $q \gets 0$;
\WHILE{$Q \neq \emptyset \wedge q<len(p)$}
	\STATE $n_i=Q.dequeue()$; $q \gets q+1$;
	\IF{$\exists n_j \in n_i.children$, such that $n_j.\mathbb{L}_j$ is equal to $q$-th item of $p$}
		\STATE $Q.inqueue(n_j)$;
	\ENDIF
\ENDWHILE
\STATE $n_p=Q.dequeue()$;
\STATE Collect edges $E_p(k)$ from $n_p.\mathbb{E}_p$ where $\forall e$, $w(e) \geq k$;
\STATE Induce $\Phi_p(k)$ from $E_p(k)$.
\RETURN $\Phi_p(k)$
\end{algorithmic}
\end{algorithm}

The query processing algorithm is outlined in Algorithm~\ref{al:query}. The general idea is that it traverses the RH-Index in a breadth-first order to locate the specific node representing the input pattern, and collects the appropriate edges from this node to form the subgraphs, i.e., the route hotspots for answering the queries. Specifically, we first initialize a queue $Q$ with the root node $n_0$ and a parameter $q$ with 0  (Line 1), where $q$ counts the number of layers that we have traversed along the RH-Index tree. The algorithm first visits a node $n_i$ of RH-Index ($n_i$ starts from $n_0$). In each step, it increments $q$ and visits the children of $n_i$ in the next layer to select the appropriate node (say $n_j$)  that follows the label sequence of $p$ (Lines 3-4), and put $n_j$ into $Q$ (Line 5). Then it continues to find the appropriate children of $n_j$. This process iterates until $Q$ is empty or $q$ reaches the length of $p$ (Line 2). If $Q$ is not empty, $n_p$ is dequeued from $Q$, which represents the pattern $p$. It then selects the edges from $\mathbb{E}_p$ whose edge weights are greater than or equal to $k$ (Line 9). Finally, the output is the connected subgraphs that are built with the selected edges (Lines 10-11).

The following example illustrates the quick query answering. Given the RH-Index in Figure~\ref{fig:thIndex}, suppose there is a query $(\langle PS, MS \rangle, 3)$. It first visits the first layer nodes and obtain the node $n_1$, where $n_1.\mathbb{L}_1=PS$. Then it visits the children of $n_1$, and obtains the node $n_6$, where $n_6.\mathbb{L}_6=MS$. From $n_6.\mathbb{E}_6$, it selects the edges whose weights are greater than or equal to $3$, and finally it recovers the route hotspots for $\langle PS, MS \rangle$ and $k=3$ based on the selected edges.
 
In summary, RH-Index is simple to construct and efficient to query. We will evaluate its effectiveness and scalability in the experimental section.

\newcommand{\parawidth}{42mm}
\section{Experiments}
\label{sec:experiments}
In this section, we evaluate the effectiveness and efficiency of our proposed algorithms on real-world networks. 

\subsection{Experimental Setup}

\subsubsection{\textbf{Evaluation Datasets}}
The following 4 real-world data sets are used for evaluation.  

\begin{table}
\caption{Graph statistics. \#Vertices, \#Edges and \#Routes indicate the number of vertices, the number of edges and the number of routes respectively. $k_{max}$ is the maximum value of $k$ for the graph. \#AT is the average length of the routes.}
\centering
\begin{adjustbox}{width=\columnwidth}
\label{tab:graphStatistics}
	\begin{tabular}{|c|c|c|c|c|c|}		
	\hline
	 & \#Vertices & \#Edges & \#Routes & \#AT & $k_{max}$ \\
	\hline
	GW & $5.0 \times 10^4$ & $1.3 \times 10^6$ & $3.5 \times 10^5$ & 6.31 & 16 \\
	\hline
	WB & $2.0 \times 10^5$ & $4.9 \times 10^6$ & $1.0 \times 10^6$ & 7.90 & 31 \\
	\hline
	CN & $1.0 \times 10^6$ & $3.9 \times 10^6$ & $5.0 \times 10^6$ & 5.03 & 58  \\
	\hline
	YELP & $1.4 \times 10^5$ & $5.6 \times 10^6$ & $4.9 \times 10^5$ & 4.00 & 61 \\
	\hline	
	\end{tabular}
	\end{adjustbox}
\end{table}

\begin{itemize}
\item \textbf{CN \cite{Tang2008}} is a collaboration network extracted from DBLP \cite{DBLP}. We regard each author as a vertex and each collaboration paper between two authors as an undirected edge. The label of a vertex represents the author's main research topic. In addition to collaborative relationships that are used to construct the network, we also consider the authors' citation relationships to build the citation sequence that is used as the route. Specifically, for every two consecutive authors A and B in the sequence, B should have cited at least one paper published by A. In CN, the detection of the route hotspot is to identify a group of closely collaborated authors that are covered by a set of routes (i.e., citing sequences), indicating they have common evolving research interests.

\item \textbf{GW \cite{Cho2011}} is a location-based dataset collected from Gowalla where users share their locations by checking-in. To construct GW, we first run k-means \cite{kmeans} to cluster proximate locations as a common vertex, and then connect two vertices if there exists a user moving between them. The labels of vertices are generated randomly as positive integers ranging from $1$ to $100$. A route represents the path of a user in chronological order. In GW, the detection can discover groups of friends sharing the same traveling routes, indicating they have the common movement and mobility pattern.

\item \textbf{WB \cite{Zhang2013}} is a microblogging network crawled from \textit{Sina Weibo}. We regard each user as a vertex and each reciprocal following relationship between two users as an undirected edge. The label of a vertex represents the location of the user. A route represents a sequence of users along the network path according to their following time. For example, for three consecutive users A, B and C in the sequence, B follows A first and then C follows B subsequently. In WB, the detection can discover groups of \textit{Sina Weibo} users who are closely connected and meanwhile have the same following behavior.

\item \textbf{YELP \cite{yelp}} is built by treating each business location as a vertex associated with a unique label. If two businesses are reviewed by at least one common user, there would be an edge between the two vertices. The routes follow the order of reviews according to their published time. In YELP, the detection can identify clusters of businesses as well as the common reviewing sequences carried by these businesses.

\end{itemize}

The network statistics are shown in Table~\ref{tab:graphStatistics}.

\subsubsection{\textbf{Comparison Methods}}
\begin{itemize}
\item \textbf{CuTS}~\cite{Jeung2008} aims to find convoys which are groups of objects traveling together for a certain period of time. As it is not designed to process graph data, to compare with our proposal, we make a minor modification, that is, we use NG-DBSCAN~\cite{Lulli2016} instead of DBSCAN~\cite{Ester1996} as the clustering algorithm. Note that we consider the convoys as the route hotspots when using CuTS.

\item \textbf{CuTSG} extends CuTS in that it first finds convoys by using CuTS, and then determines whether a convoy can also cover a connected maximal $k$-truss with as many vertices, edges and routes as possible.

\item \textbf{GreedyRH} is the greedy algorithm described in Algorithm~\ref{al:simpleAl}.

\item \textbf{FastRH} is our core algorithm described in Algorithm~\ref{al:thd}. We apply both the properties of pattern anti-monotonicity and hotspot anti-monotonicity on GreedyRH.

\item \textbf{PFastRH} is the parallel version of FastRH, which is implemented by using multithreading.

\item \textbf{GreedyRHP} is an improved algorithm based on GreedyRH, where we apply the property of pattern anti-monotonicity on GreedyRH.

\item \textbf{GreedyRHH} is the other improved algorithm based on GreedyRH, where we apply the property of hotspot anti-monotonicity on GreedyRH.
\end{itemize}

\subsubsection{\textbf{Evaluation Metrics}}

\begin{itemize}
\item \textbf{Precision Score (PS)}. $\mathcal{C}$ denotes the complete set of hotspots obtained by GreedyRH and $\hat{\mathcal{C}}$ denotes the set of hotspots obtained by other methods, such as FastRH, GreedyRHP, GreedyRHH and PFastRH. PS is computed as $PS=\frac{|\mathcal{C} \cap \hat{\mathcal{C}}|}{|\mathcal{C}|}$. This metric is used to verify our proposal works correctly with the anti-monotonicity and independence properties (i.e., Theorem~\ref{th:pattern}, Theorem~\ref{th:graph} and Proposition~\ref{prop:distributed}). Please note that, as CuTS and CuTSG don't use these properties to prune unqualified route hotspots or achieve parallel speedup, they are not evaluated in terms of PS.

\item \textbf{Time Cost and Space Cost}. ``Time Cost" is the running time for the detection or indexing process. ``Space Cost'' is the main memory used for indexing.

\item \textbf{Number of Patterns (\#NP) and Number of Hotspots (\#NH)}. ``Number of Patterns" and ``Number of Hotspots" refer to the total number of patterns that are used to support hotspots and the total number of hotspots that are obtained with our methods respectively. We use \#CuNP, \#CuGNP and \#RHNP (resp. \#CuNH, \#CuGNH and \#RHNH) to represent the total number of patterns (resp. number of hotspots) obtained with CuTS, CuTSG and PFastRH.
\end{itemize}

\subsubsection{\textbf{Implementation Details}}
All algorithms are implemented in Java and all of the experiments are conducted on a commodity PC with Linux 16.04, Core-i7 6700K CPU (4.00GHz) and 64 GB main memory. We use 8 threads for running PFastRH. To evaluate the impacts of $min\_sup$, we set $min\_sup=100, 200, 300, 400, 500$ for CN, and $min\_sup=10, 20, 30, 40, 50$ for GW, WB and YELP respectively.

\subsection{Experimental results}
\subsubsection{\textbf{Effectiveness Analysis}}
We evaluate the effectiveness of these methods on WB, GW, CN and YELP, but omit the results of GreedyRH as it is too slow on large datasets.

\begin{table}
\caption{The PS on the GW, WB, YELP and CN.}
\centering
\begin{adjustbox}{width=\columnwidth}
\label{tab:ps}
	\begin{tabular}{|c|c|c|c|c|c|}
	\hline
	\multicolumn{6}{|c|}{GW/WB/YELP}  \\
	\hline
	$min\_sup$  & 10 & 20  & 30  & 40 & 50  \\
	\hline
	PS (FastRH)   & $100\%$ & $100\%$  & $100\%$  & $100\%$ & $100\%$  \\
	\hline
	PS (GreedyRHP)   & $100\%$ & $100\%$  & $100\%$  & $100\%$ & $100\%$  \\
	\hline
	PS (GreedyRHH)   & $100\%$ & $100\%$  & $100\%$  & $100\%$ & $100\%$  \\
	\hline
	PS (PFastRH)   & $100\%$ & $100\%$  & $100\%$  & $100\%$ & $100\%$  \\
	\hline
	\multicolumn{6}{|c|}{CN}  \\
	\hline
	$min\_sup$  & 100 & 200  & 300  & 400 & 500  \\
	\hline
	PS (FastRH)   & $100\%$ & $100\%$  & $100\%$  & $100\%$ & $100\%$  \\
	\hline
	PS (GreedyRHP)   & $100\%$ & $100\%$  & $100\%$  & $100\%$ & $100\%$  \\
	\hline
	PS (GreedyRHH)   & $100\%$ & $100\%$  & $100\%$  & $100\%$ & $100\%$  \\
	\hline
	PS (PFastRH)   & $100\%$ & $100\%$  & $100\%$  & $100\%$ & $100\%$ \\
	\hline
	\end{tabular}
\end{adjustbox}
\end{table}

Table~\ref{tab:ps} shows the precision scores with varying $min\_sup$ for all compared methods. We can see that the PS values of FastRH, GreedyRHP, GreedyRHH, and PFastRH on GW, WB, CN, and YELP are always equal to $100\%$. It indicates that, FastRH, GreedyRHP, GreedyRHH, and PFastRH won't change the number of hotspots, which validates the correctness of our pattern and hotspot anti-monotonicity properties as well as the independent properties.

\subsubsection{\textbf{Analysis on Patterns and Route Hotspots}}
\begin{figure}[htb]
\centering	
	\subfigure[\#NH and \#NP (GW)]{\includegraphics[width=\parawidth]{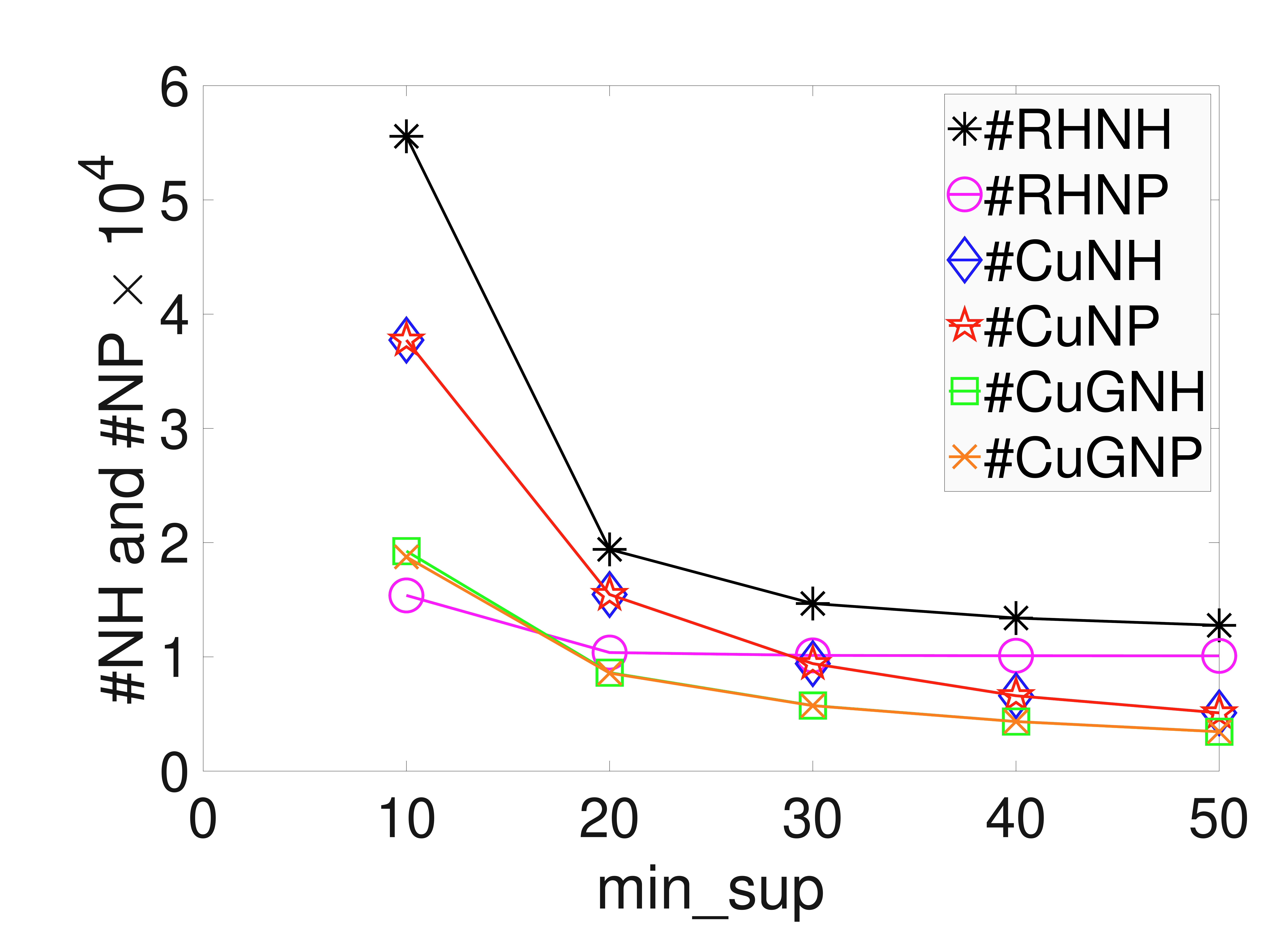}}	
	\subfigure[\#NH and \#NP (WB)]{\includegraphics[width=\parawidth]{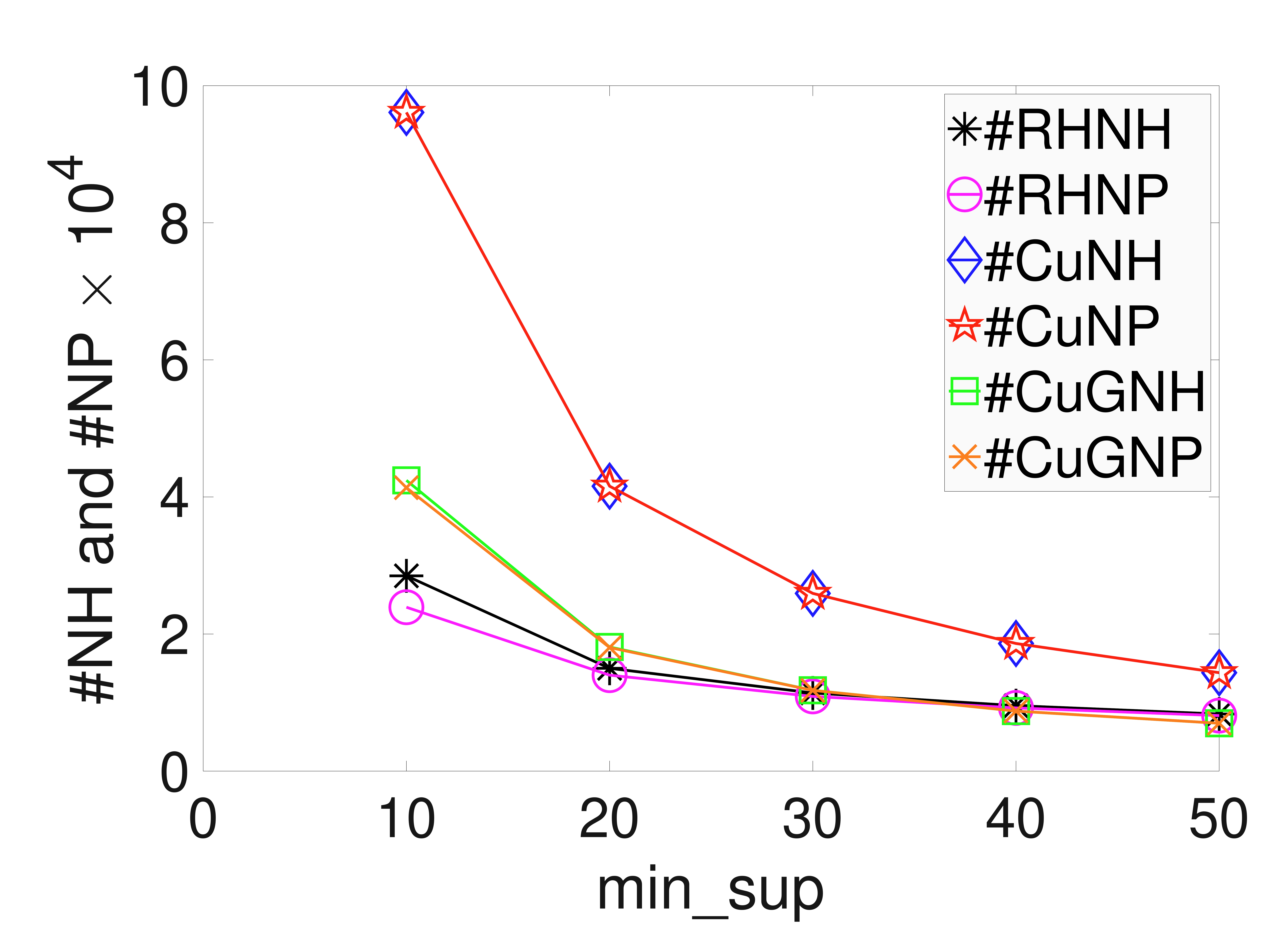}}	
		
	\subfigure[\#NH and \#NP (YELP)]{\includegraphics[width=\parawidth]{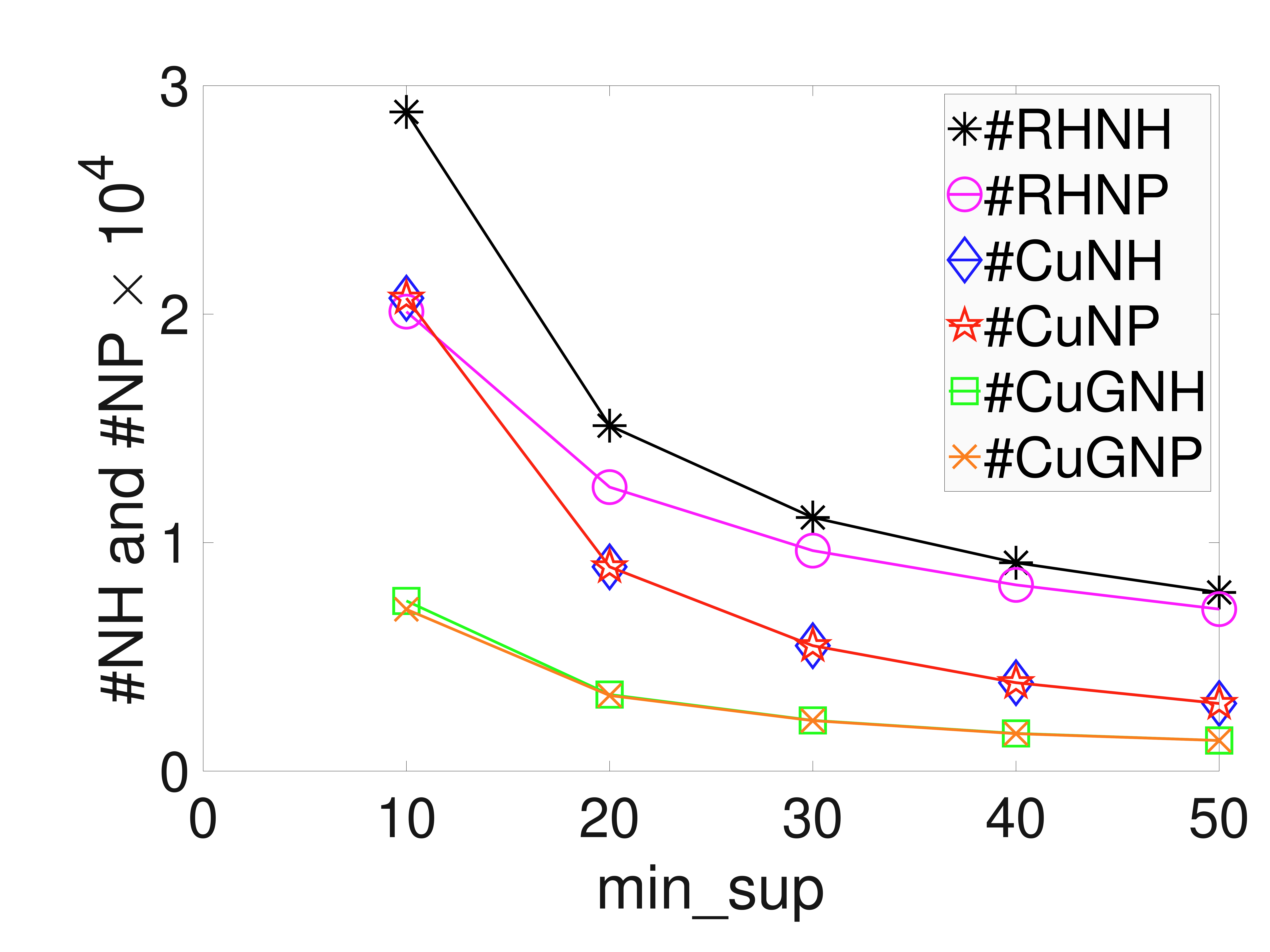}}				
	\subfigure[\#NH and \#NP (CN)]{\includegraphics[width=\parawidth]{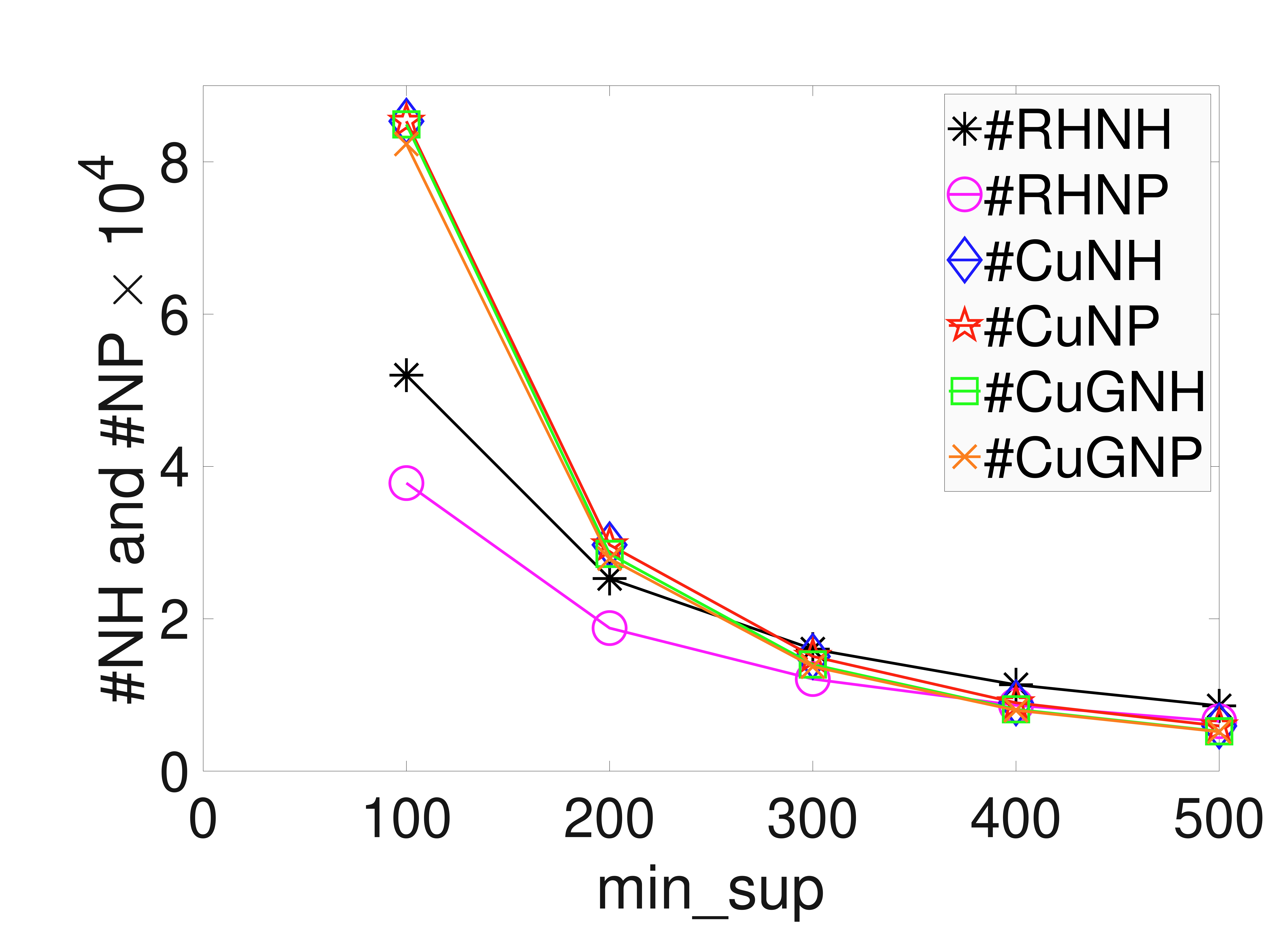}}	
\caption{\#NH and \#NP on the GW, WB, YELP, and CN.}
\label{fig:npnh} 
\end{figure}

We analyze the total number of patterns and route hotspots obtained with PFastRH, CuTS and CuTSG by varying $min\_sup$, and the results are shown in Figure~\ref{fig:npnh}.

It can be observed that \#NP and \#NH decrease when $min\_sup$ increases. Generally, \#NH would be greater than or equal to \#NP as one pattern may correspond to multiple hotspots, but we can observe in Figure~\ref{fig:npnh} that \#NH is almost equal to \#NP when $min\_sup$ is large. The reason is that larger $min\_sup$ may make some route hotspots become unqualified as their numbers of routes are now less than $min\_sup$. Note that, \#CuNH is equal to \#CuNP, as CuTS considers a pattern (i.e., a convoy) as a hotspot.

Also, as shown in Figure~\ref{fig:npnh}, \#CuNP and \#CuNH are larger than \#CuGNP and \#CuGNH, because the definition of $k$-truss-based convoys defined in CuTSG is more rigorous than that defined in CuTS. In the datasets WB and CN, \#CuNP and \#CuGNP (resp. \#CuNH and \#CuGNH) are greater than \#RHNP (resp. \#RHNH) when $min\_sup$ is small. This is because the definition of route hotspot defined in Definition~\ref{def:communityDef} uses both pattern and graph information, which is able to avoid a large amount of unqualified patterns. However, when $min\_sup$ increases, \#CuNP and \#CuGNP become smaller than \#RHNP, and \#CuNH and \#CuGNH are also smaller than \#RHNH. This is because CuTS and CuTSG only use route information but ignore the sequential patterns in the routes. In the datasets GW and YELP, \#RHNH is greater than \#CuNH and \#CuGNH with different $min\_sup$. We can also observe that, for each dataset, the gap between \#RHNH and \#RHHP is larger than that between \#CuGNH and \#CuGNP, as well as that between \#CuNH and \#CuNP. The reason is that, our method can find more densely interconnected subgraphs with larger values of $k$.

\subsubsection{\textbf{Efficiency of Hotspot Finding}}
\begin{figure}[htb]
\centering
	\subfigure[Time Cost (GW)]{\includegraphics[width=\parawidth]{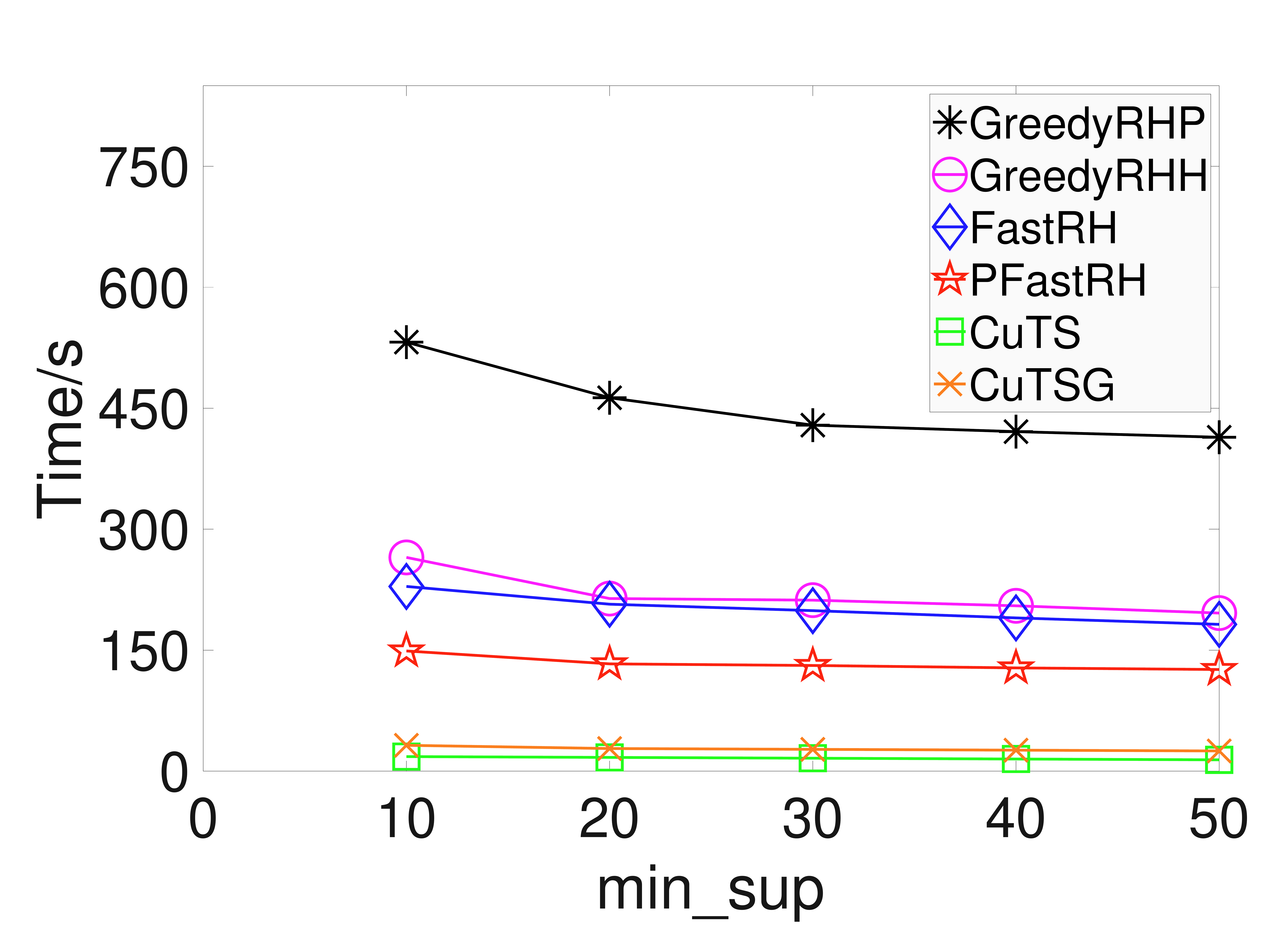}}
	\subfigure[Time Cost (WB)]{\includegraphics[width=\parawidth]{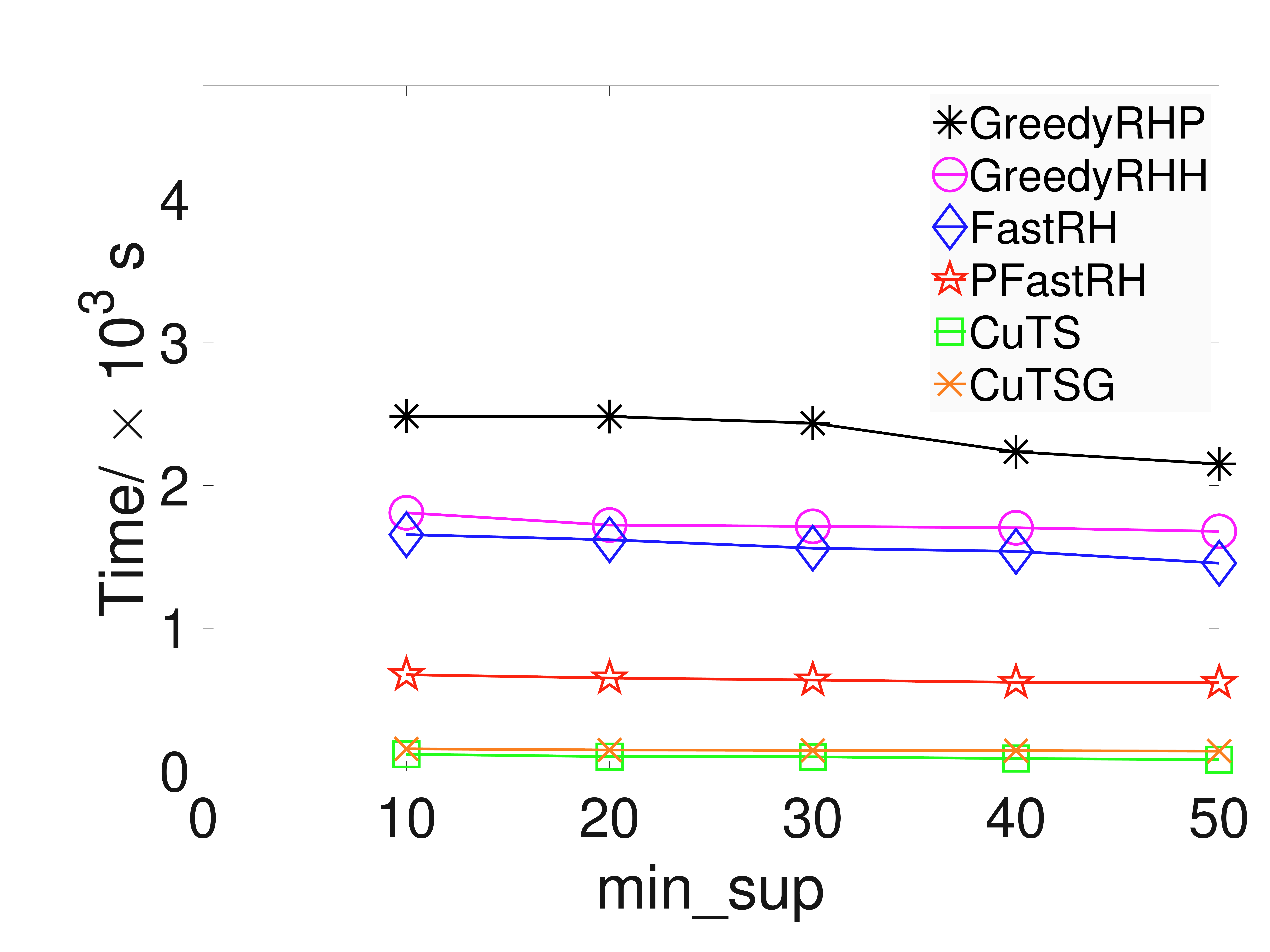}}	
	
	\subfigure[Time Cost (YELP)]{\includegraphics[width=\parawidth]{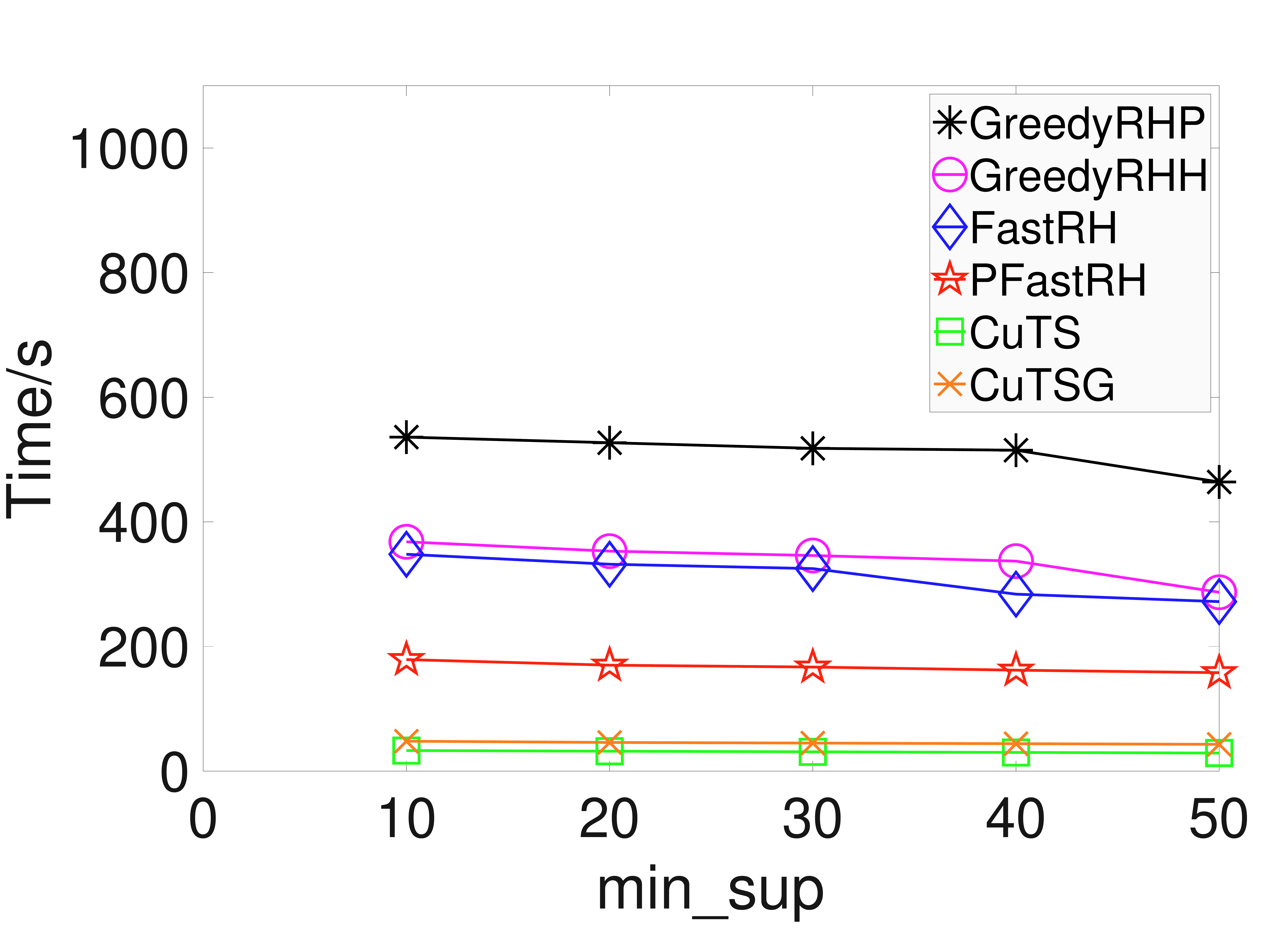}}	
	\subfigure[Time Cost (CN)]{\includegraphics[width=\parawidth]{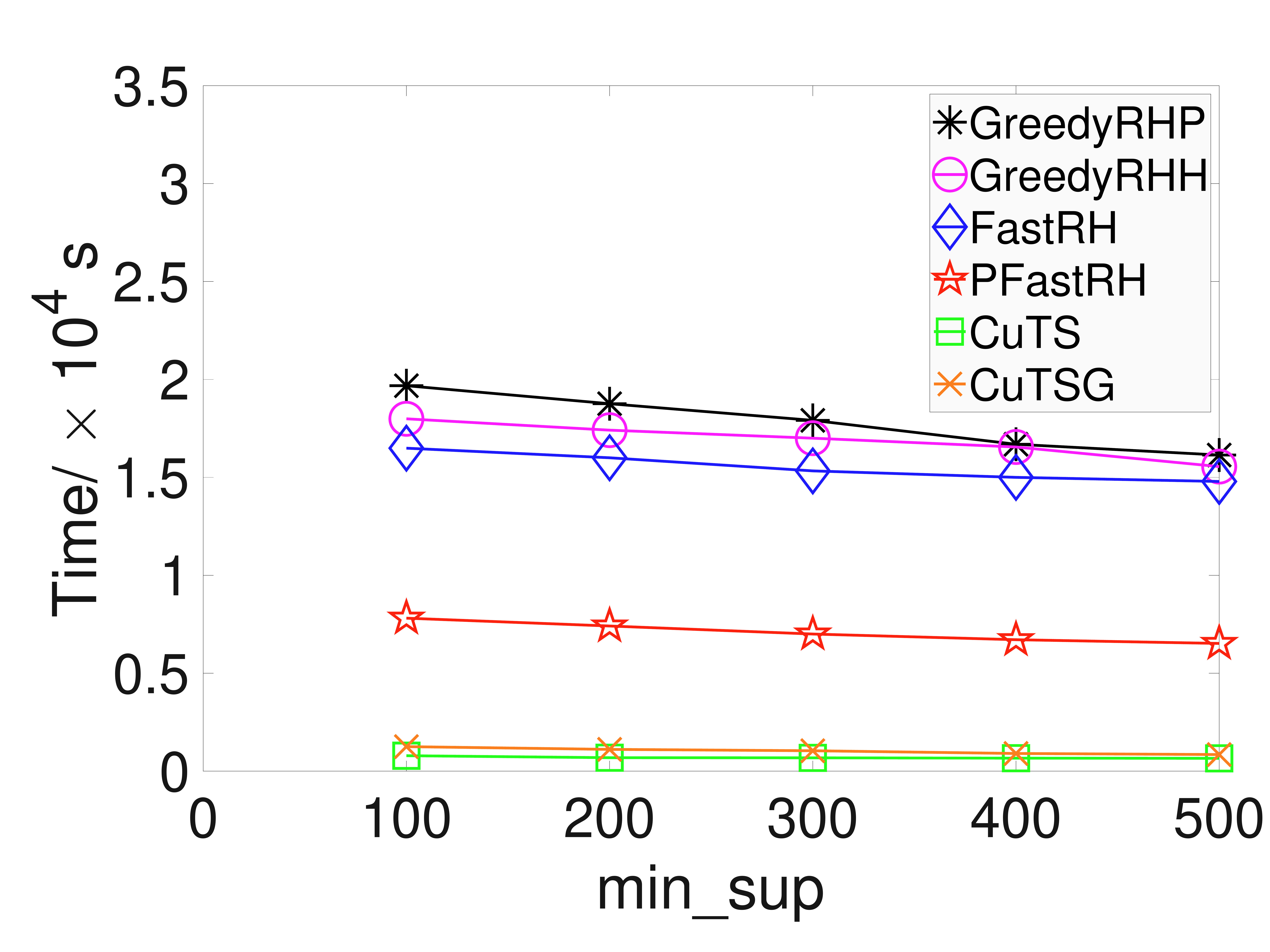}}			
\caption{The Time Cost on the GW, WB, YELP, and CN.}
\label{fig:runTime} 
\end{figure}

We evaluate the efficiency of FastRH, GreedyRHP, GreedyRHH, PFastRH, CuTS and CuTSG by varying $min\_sup$. The performances in terms of time cost are shown in Figure~\ref{fig:runTime}, where the time cost is the running time for a specific algorithm. Please note that as GreedyRH is too slow, we omit its results.

As shown in Figure~\ref{fig:runTime}, when $min\_sup$ increases, the time cost decreases  due to the decrease of \#NP and \#NH shown in Figure~\ref{fig:npnh}. We can observe that GreedyRHH is more efficient than GreedyRHP. The reason is that GreedyRHH uses the hotspot anti-monotonicity to prune unnecessary candidate values of $k$, and this pruning policy can prevent a significant number of checking operations. This is validated in Table~\ref{tab:kStat} that the number of route hotspots drops sharply with the increase of $k$. FastRH is faster than GreedyRHH and GreedyRHP, since FastRH uses two pruning rules to speed up the hotspot finding process. PFastRH is even faster than FastRH because PFastRH is a parallel method. CuTS is the fastest one since it only uses the route information. CuTSG is faster than GreedyRH and its variations, as it doesn't consider the patterns in routes and thus avoids the sequential pattern mining process. Please note that another reason why CuTS and CuTSG are faster than PFastRH is that we only use 8 threads in our experiments (according to our machine), which limits the speed-up ratio of the parallelized algorithm.

\begin{figure}[htb]
\centering
	\subfigure[\#NP with \#Vertices]{\includegraphics[width=\parawidth]{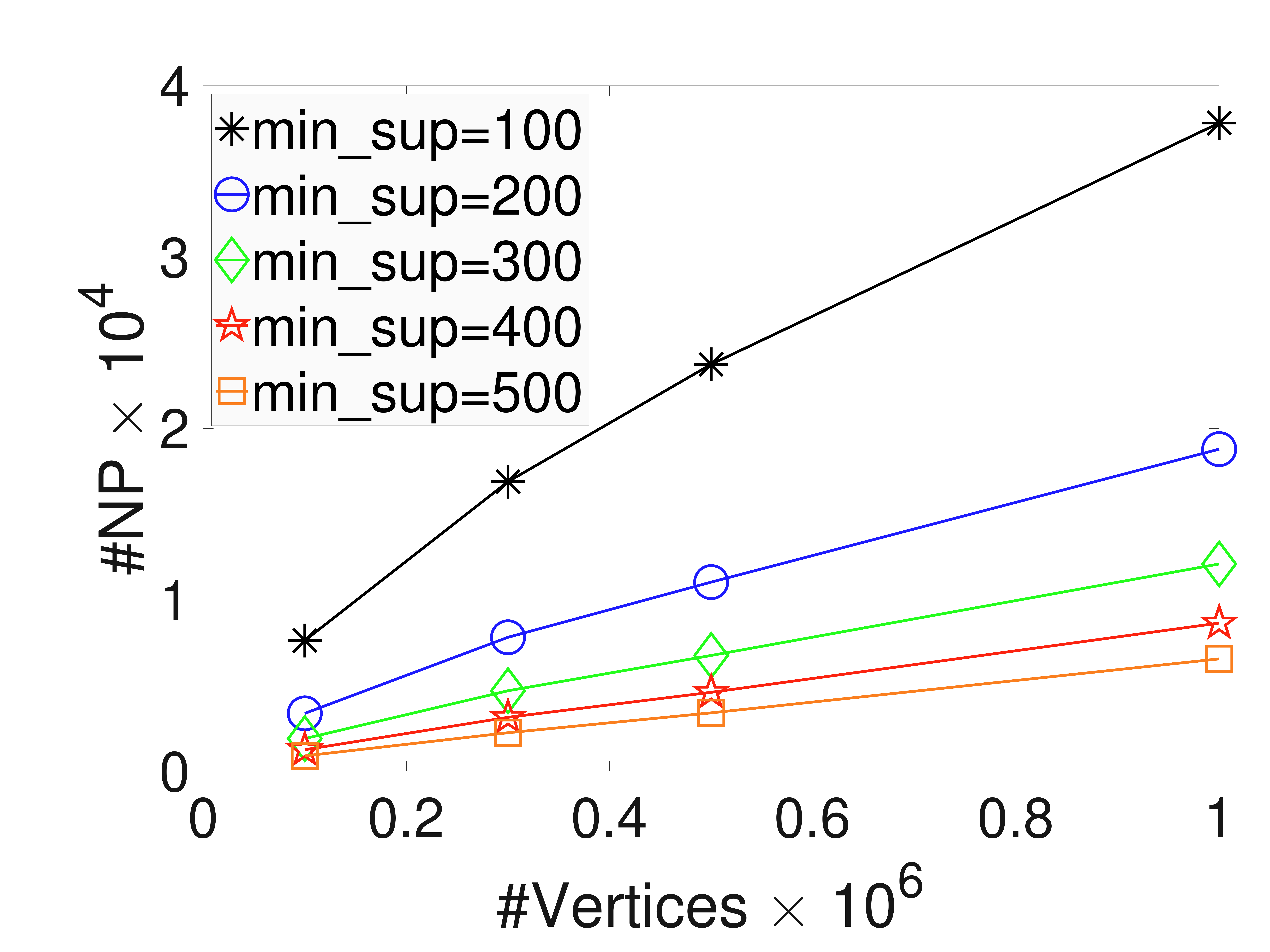}}
	\subfigure[\#NP with \#Routes]{\includegraphics[width=\parawidth]{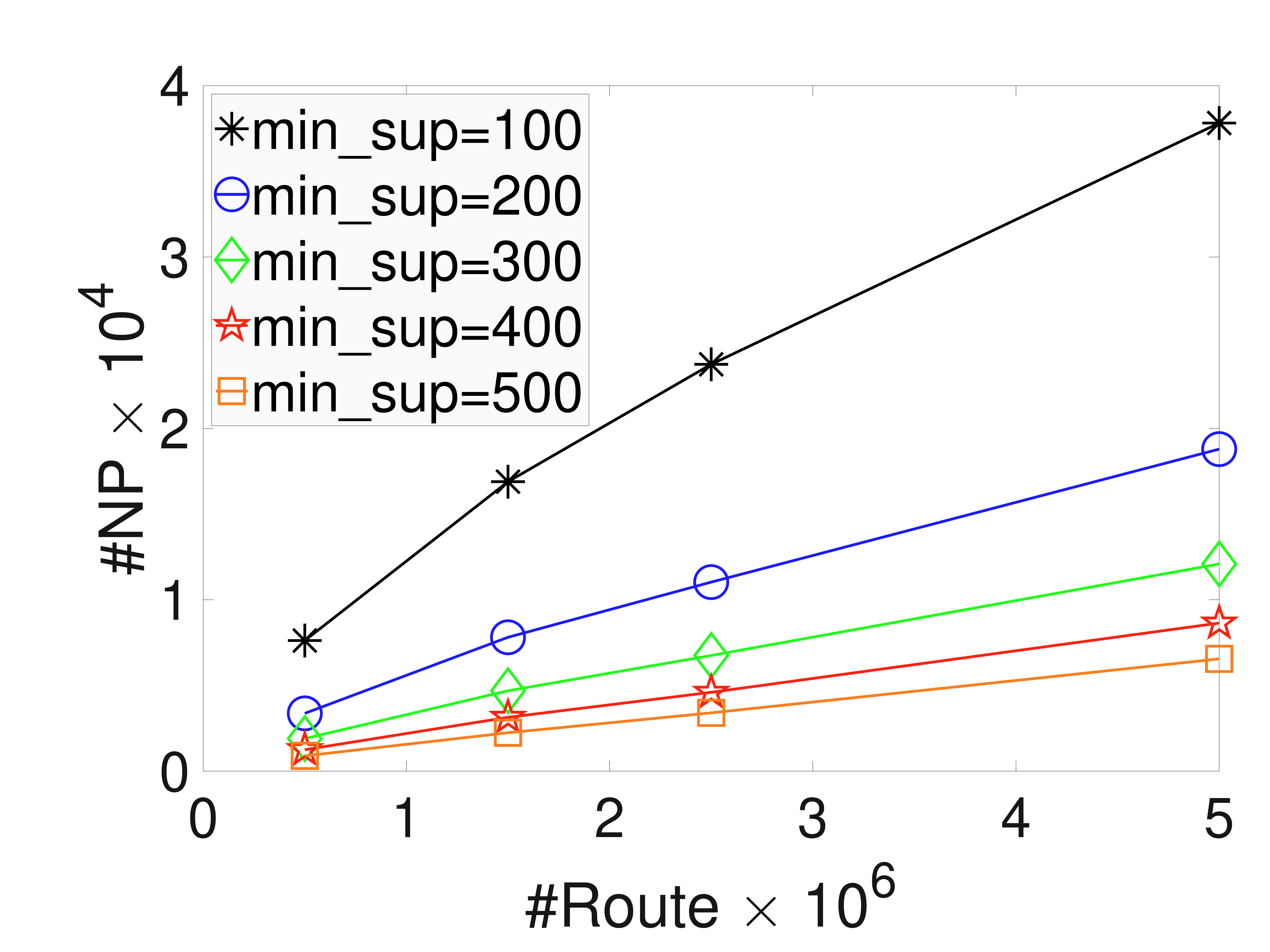}}
	
	\subfigure[\#NH with \#Vertices]{\includegraphics[width=\parawidth]{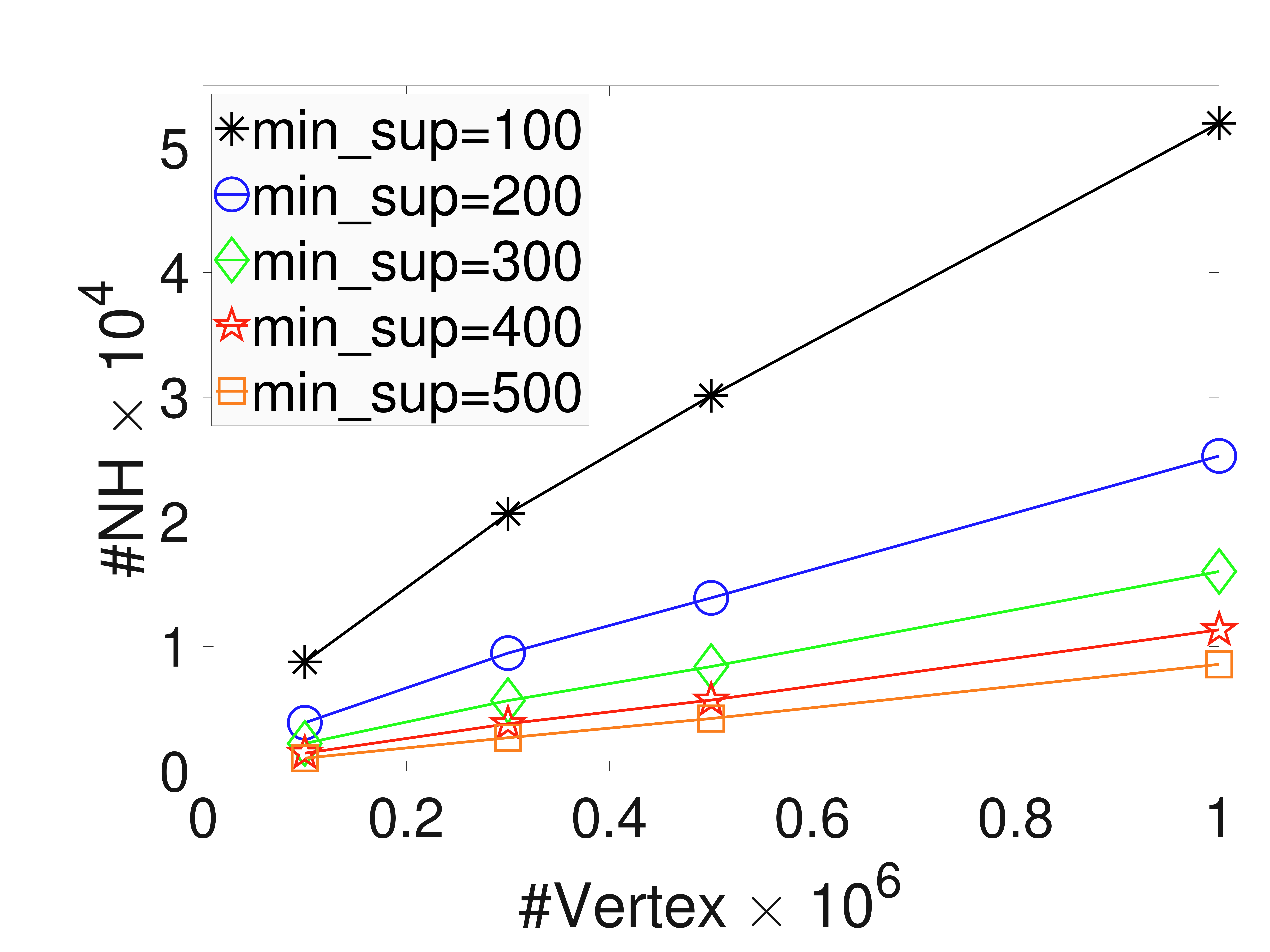}}
	\subfigure[\#NH with \#Routes]{\includegraphics[width=\parawidth]{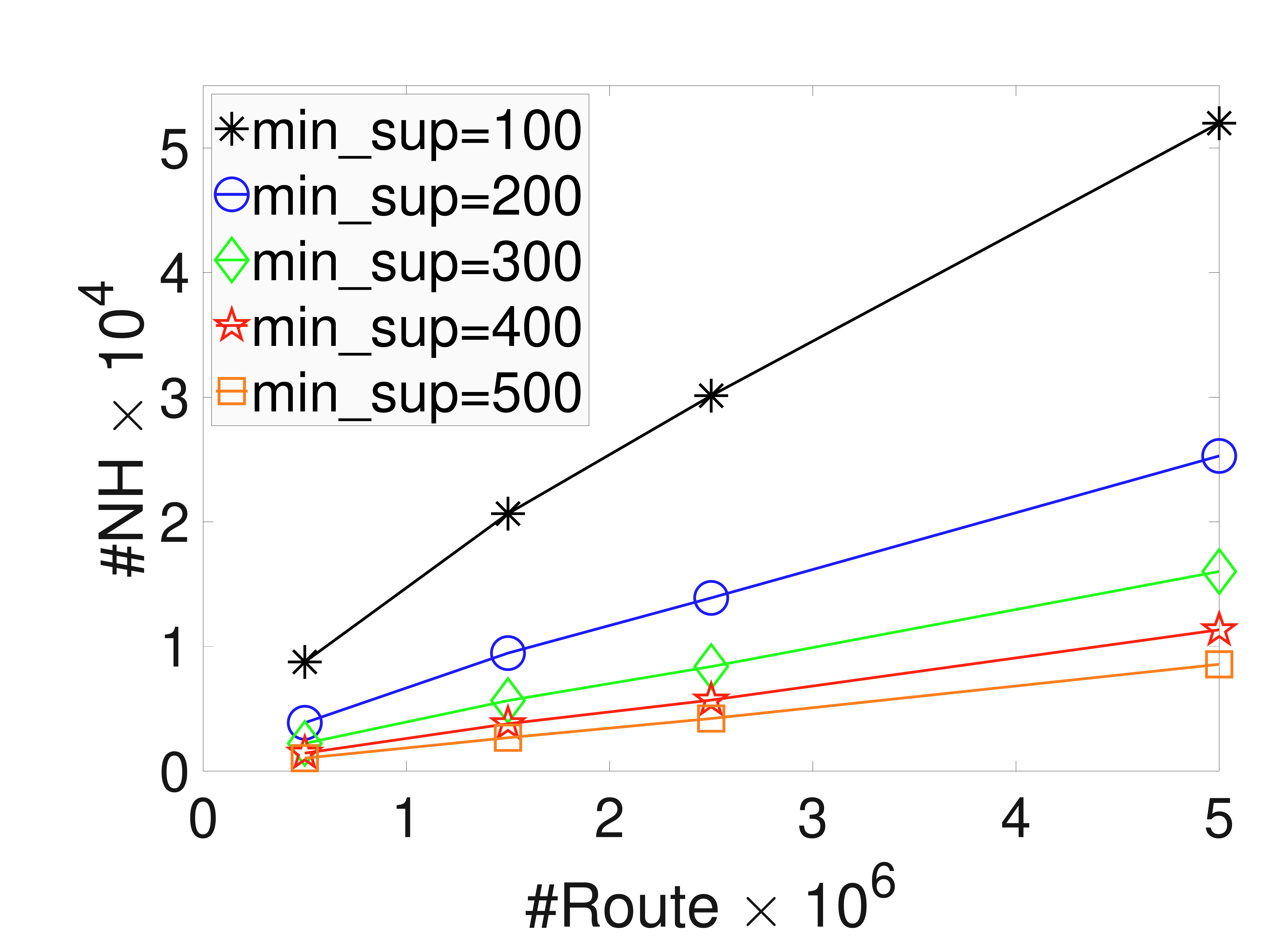}}	
	
	\subfigure[Time Cost with \#Vertices]{\includegraphics[width=\parawidth]{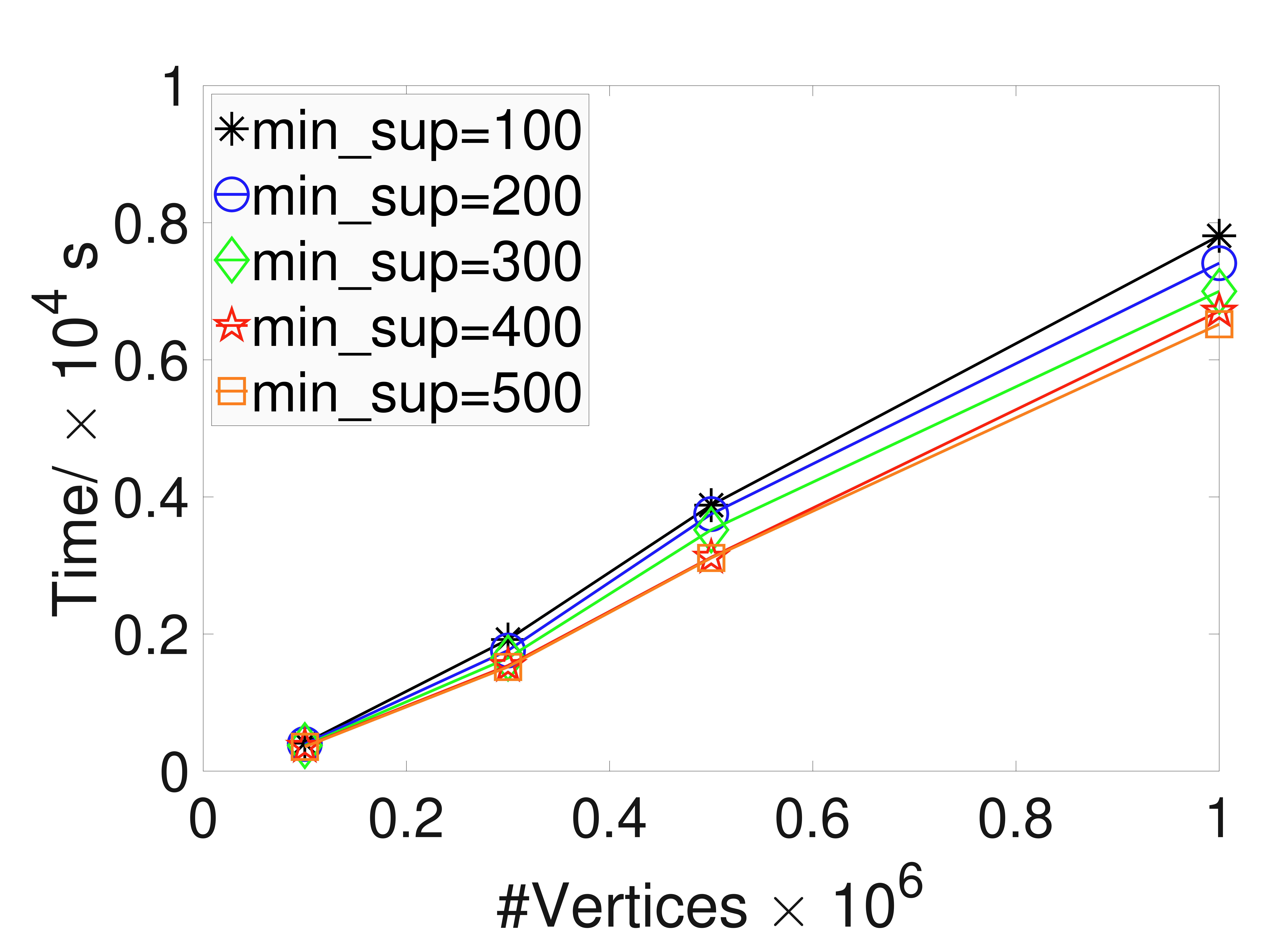}}
	\subfigure[Time Cost with \#Routes]{\includegraphics[width=\parawidth]{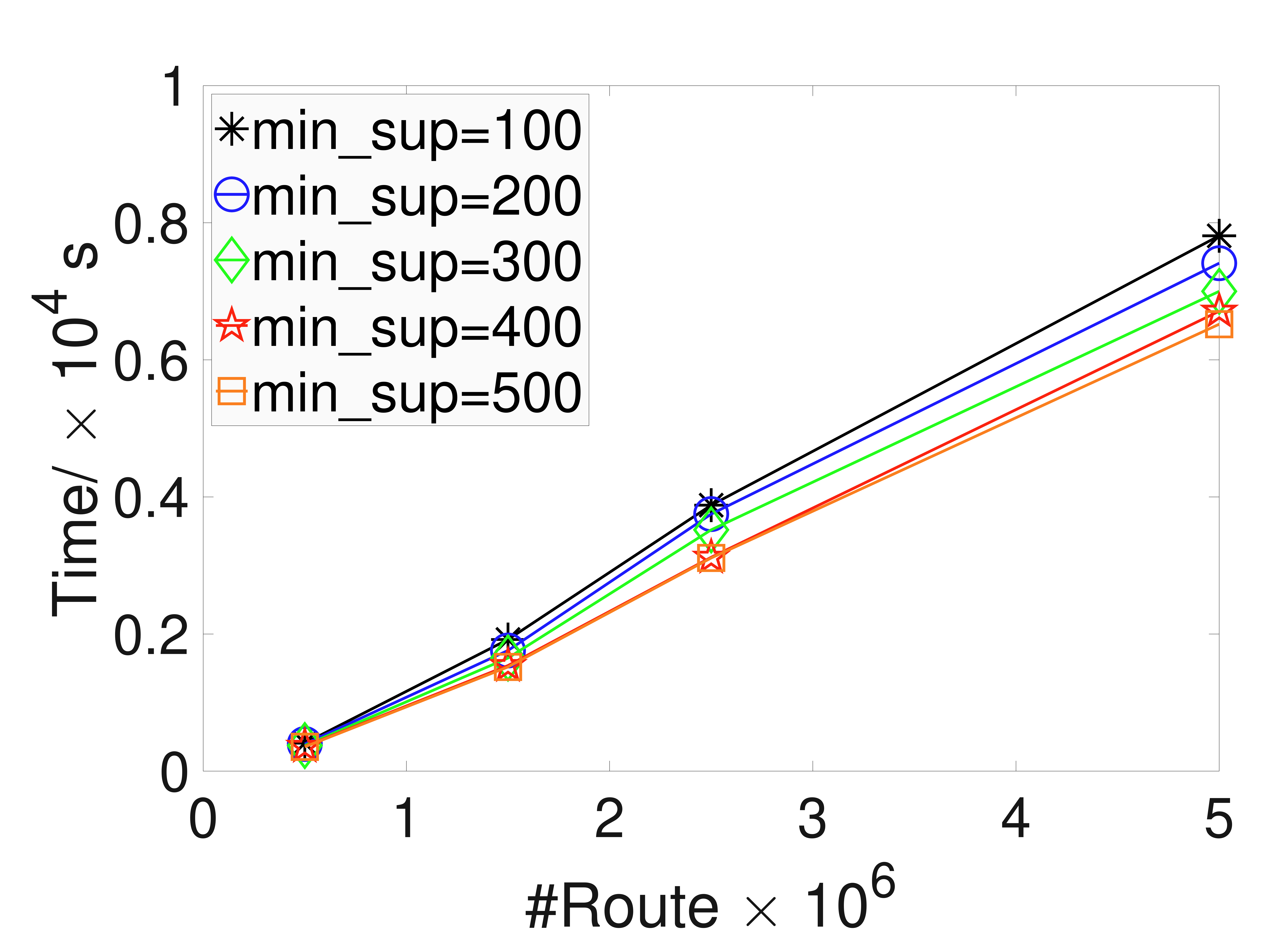}}	
	
\caption{The \#NP, \#NH and Time Cost on sampled subgraphs of CN with different values of $min\_sup$, where \#Vertices=$1.0 \times 10^5$, $3.0 \times 10^5$, $5.0 \times 10^5$, $1.0 \times 10^6$ and \#Routes=$5.0 \times 10^5$, $1.5 \times 10^6$, $2.5 \times 10^6$, $5.0 \times 10^6$ respectively. The Time Cost is obtained by PFastRH.}
\label{fig:cn} 
\end{figure}
 
As shown in Figure~\ref{fig:cn}, with the increasing number of vertices and routes for CN, the time cost increases almost linearly. This is because a greater number of vertices and routes bring in a greater number of patterns and route hotspots, i.e., \#NP and \#NH, and thus it would take longer time. The near-linear increasing demonstrates that the distribution of route hotspots in the network is near-uniform.

\subsubsection{\textbf{Indexing and Query Processing}}
\begin{table}
\caption{An Example of NRH-Index.}
\centering
\begin{adjustbox}{max width=\columnwidth}
\label{tab:nth}
	\begin{tabular}{|c|c|c|c|}
	\hline
	ID & Pattern & $k$ & Edges \\
	\hline
	1 & $\langle a, b \rangle$ & 2 & $(v_1, v_2)$, $(v_1, v_3)$, $(v_2, v_3)$, $(v_2, v_4)$, $(v_2, v_5)$, $(v_3, v_4)$, $(v_3, v_5)$ \\
	\hline
	2 & $\langle a, b \rangle$ & 3 & $(v_1, v_2)$, $(v_1, v_3)$, $(v_2, v_3)$, $(v_2, v_4)$, $(v_2, v_5)$, $(v_3, v_4)$, $(v_3, v_5)$  \\
	\hline
	$\ldots$ & $\ldots$ & $\ldots$ & $\ldots$ \\
	\hline
	\end{tabular}
\end{adjustbox}
\end{table}

We evaluate the scalability and the efficiency of RH-Index. Since it's trivial for the situation when a hotspot owns just one route, we fix $min\_sup=2$. Please note that a larger $min\_sup$ can result in smaller RH-Index. For comparing with RH-Index, we propose a na\"ive method called \textit{NRH-Index}, where each hotspot is stored as a set of edges in the row of a big table. NRH-Index can also support querying by patterns via traversing rows in that big table. Table~\ref{tab:nth} shows an example of NRH-Index with pattern $\langle a, b \rangle$ and $k=2, 3$. Please note that the IDs in this example are assigned randomly.

\begin{table}
\caption{Index statistics (Size in Megabytes and Time in seconds)}
\centering
\begin{adjustbox}{width=\columnwidth}
\label{tab:indexStatistics}
	\begin{tabular}{|c|c|c|c|c|}
	\hline
	 & GW & WB  & YELP  & CN \\
	\hline
	\#MNRH & 47,924 & 44,035 & 27,102  & 39,918 \\
	\hline
	\#MRH & 13,336 & 28,181 & 19,776 &  15,403  \\
	\hline
	\#TNRH & 382 & 865 & 945  & 8,392 \\
	\hline
	\#TRH & 294 & 691 & 934 & 8,166 \\	
	\hline
	\#NP & $8.6 \times 10^5$ & $1.6 \times 10^5$ & $1.8 \times 10^5$ &  $6.7 \times 10^5$ \\
	\hline
	\#NH & $2.1 \times 10^6$ & $3.8 \times 10^6$ & $5.4 \times 10^5$ & $8.2 \times 10^5$ \\
	\hline
	\#AVH & 20.6 & 95.5 & 42.6 & 15.5 \\
	\hline
	\#AEH & 25.0 & 131.2 & 51.3 &  17.6 \\
	\hline
	\#ARH & 4.0 & 19.7 & 9.5 & 32.2 \\
	\hline
	\#ALH & 6.31 & 2.82 & 4.00 &  3.39 \\
	\hline
	$max\{k\}$ & 6 & 7 & 20 & 12 \\
	\hline
	\end{tabular}
\end{adjustbox}
\end{table}

The indexing performance of RH-Index is shown in Table~\ref{tab:indexStatistics}, where ``\#MNRH'' and ``\#MRH'' are the cost of the main memory for NRH-Index and RH-Index respectively; ``\#TNRH'' and ``\#TRH" denote the time cost to build NRH-Index and RH-Index respectively, including both detection time and indexing time; ``\#AVH'', ``\#AEH'', ``\#ARH'', ``\#ALH'' are the average number of vertices, edges, routes and the length of patterns for each hotspot. As shown in Table~\ref{tab:indexStatistics}, over all the datasets, the memory cost of NRH-Index (i.e., \#MNRH) is much larger than that of RH-Index (i.e., \#MRH), and similarly, the time cost of NRH-Index (i.e., \#TNRH) is  also greater than that of RH-Index (i.e., \#TRH). The main reason is that RH-Index uses a tree-based structure to index the route hotspots, which can save both space and time cost compared with the table-based structure.

\begin{table}
\caption{\#NH for different values of $k$ on all datasets}
\centering
\begin{adjustbox}{width=\columnwidth}
\label{tab:kStat}
	\begin{tabular}{|c|c|c|c|c|}
	\hline
	& GW & WB  & YELP  & CN \\
	\hline
	$k=2$ & $1.7 \times 10^6$ & $3.6 \times 10^5$ & $5.2 \times 10^5$ &  $7.8 \times 10^5$ \\
	\hline
	$k=3$ & $4.6 \times 10^5$ & $1.6 \times 10^5$ & $1.6 \times 10^4$ &  $3.1 \times 10^4$ \\
	\hline
	$k=4$ & $1.6 \times 10^4$ & $76$ & 118 &  $6,163$ \\
	\hline
	$k=5$ & 370  & 11 & 29 &  $1,707$ \\
	\hline
	$k=6$ & 9 & 4 & 18 &  602 \\
	\hline
	\end{tabular}
\end{adjustbox}
\end{table}

Table~\ref{tab:kStat} shows how $k$ affects the performance on different datasets. Since the evaluated RH-Index is obtained by fixing $min\_sup=2$, one can obtain the maximum number of hotspots when $k=2$, and this number decreases when $k$ increases. The maximum value of $k$ is 12 for CN. It can also be observed that the number of hotspots drops quickly when $k$ increases, which is beneficial for GreedyRHH.

\begin{table}
\caption{Query time(In milliseconds)}
\centering
\begin{adjustbox}{width=\columnwidth}
\label{tab:indexQuery}
	\begin{tabular}{|c|c|c|c|c|c|c|}
	\hline
	 DataSet & GW & WB  & YELP & CN1 & CN2 & CN\\
	\hline
	With NRH-Index & 0.098 & 9.468 & 0.424 & 1.789 &  3.786 & 4.715 \\
	\hline
	With RH-Index  & 0.043  & 0.864 & 0.224 & 0.067  & 0.090 & 0.122 \\
	\hline
	\end{tabular}
\end{adjustbox}
\end{table}

We compare the average query time of RH-Index and NRH-Index in Table~\ref{tab:indexQuery}. The query time includes the time to find patterns and the time to recover hotspots. It is shown that over all datasets, the time cost for querying RH-Index is much lower than that for NRH-Index, indicating that RH-Index is much more efficient even with far less memory requirement. In addition, the query time generally increases with the size of the networks.

\subsection{Case Study}
Two case studies from CN and YELP are given to explain the effectiveness of our method. 

\begin{figure}[!ht]
\centering
	\includegraphics[width=80mm]{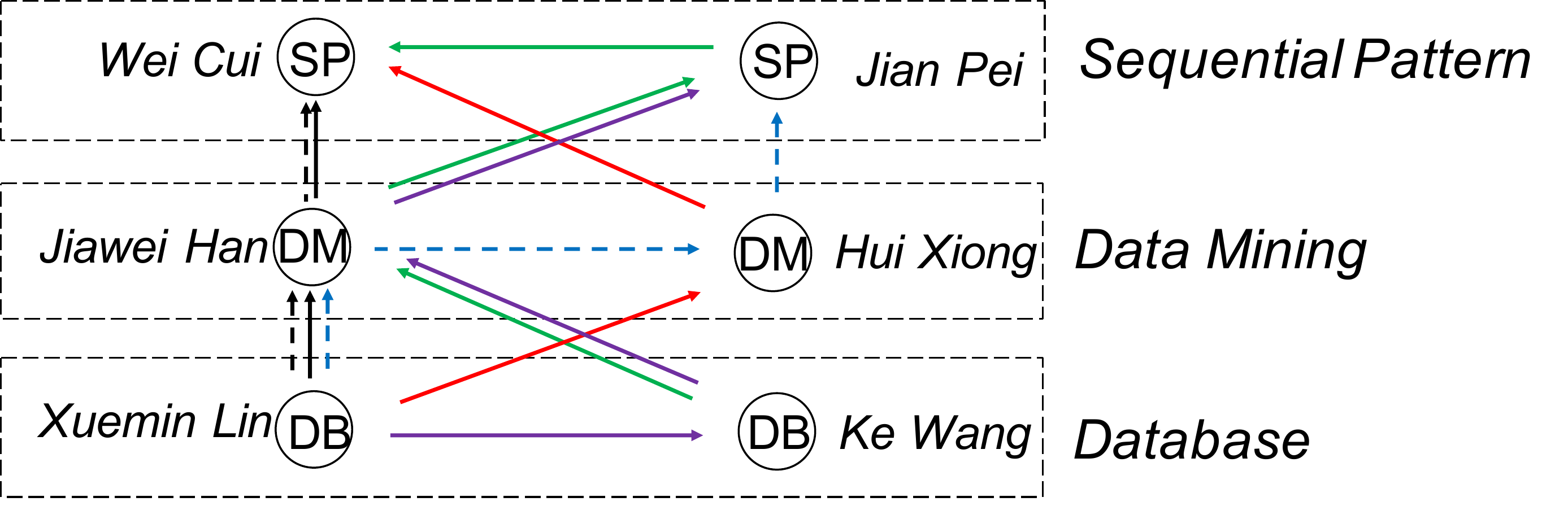}
\caption{Case Study for $\langle$DB, DM, SP$\rangle$ and $k=3$ in CN, where the dashed rectangles represent the research areas and the colored and dashed lines represent routes in the hotspot. We omit the edges.}
\label{fig:cnPattern} 
\end{figure}

Figure~\ref{fig:cnPattern} shows a route hotspot for pattern $\langle$DB, DM, SP$\rangle$ and $k=3$, where DB, DM and SP stand for Database, Data Mining and Sequential Pattern respectively. This hotspot shows how the research highlights evolved in the 2000s. During the time, the techniques for database were applied to data mining, and then applied to sequential pattern mining, and the detected scholars in theses domains had close cooperations. 

\begin{figure}[!ht]
\centering
	\includegraphics[width=70mm]{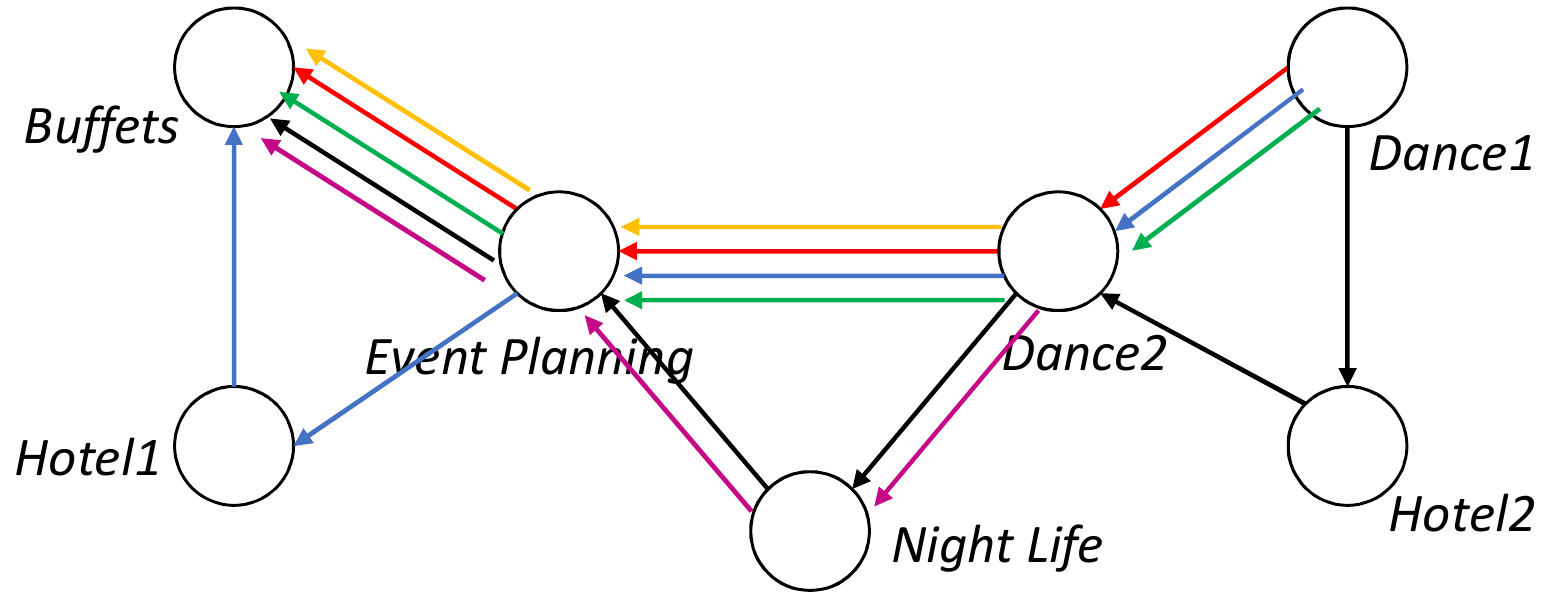}
\caption{Case Study for $\langle$Dance, Event Planning, Buffets$\rangle$ and $k=3$ in YELP, where the colored lines represent routes in the hotspot. We omit the edges.}
\label{fig:yelpPattern} 
\end{figure}

Figure~\ref{fig:yelpPattern} shows interesting relationships among dance, event planning, and buffets. Dance1 and Dance2 denote two different vertices  but have the same attribute dance, and so is for Hotel1 and Hotel2 affiliated with the same attribute Hotel. From the observed sequential pattern $\langle$Dance, Event Planning, Buffets$\rangle$, we can know that dance, event planning and buffets are tightly coupled activities, and people who participate these activities usually first dance, then followed by the planning on the subsequent activities, and finally go on for dinner. Note that, vertices whose attributes don't exist in this pattern may also appear in the hotspot (e.g., attributes Hotel and Night Life don't appear in the pattern but appear in the hotspot). The reason is that our definition is relaxed in the attributes so that triangles can be formed easily.

In contrast to the route hotspot for $\langle$Dance, Event Planning, Buffets$\rangle$ and $k=3$, as shown in Figure~\ref{fig:yelpPattern}, we cannot find the route hotspot for $\langle$Dance, Buffets, Event Planning$\rangle$ and $k=3$. The reason is that, people don't exhibit such kind of behaviors on YELP. 

\section{Conclusion}
\label{sec:con}
In this paper, we study the problem of \textit{finding route hotspots in large labeled networks}. We introduce the definition of the route hotspot and propose a novel approach \textit{FastRH} to find them, based on pattern anti-monotonicity and hotspot anti-monotonicity properties. To preserve the detected hotspots and support fast user querying by patterns, we design a novel index, called \textit{RH-Index}. Extensive experiments performed on large real-world networks demonstrate the effectiveness and efficiency of the proposed methods.

\ifCLASSOPTIONcompsoc
  \section*{Acknowledgments}
\else
  \section*{Acknowledgment}
\fi

This work is supported in part by the National Key Research and Development Program of China under Grant 2017YFB0803301, in part by Natural Science Foundation of China under Grant 61976026 and U1836215, in part by 111 Project under Grant B18008, and in part by NSF under grants III-1526499, III-1763325, III-1909323, and CNS-1930941.

\ifCLASSOPTIONcaptionsoff
  \newpage
\fi

\bibliographystyle{IEEEtran}
\bibliography{reference}
\end{document}